\documentclass[]{aa}
\usepackage{natbib}
\usepackage{graphicx}
\usepackage{color}
\usepackage{amsmath}
\usepackage{url}

\bibpunct{(}{)}{;}{a}{}{,} 

\newcommand\bb[1] {   \mbox{\boldmath{$#1$}}  }

\newcommand{\mean}[1]{\langle #1 \rangle}

\definecolor{brown}{rgb}{0.42,0.24,0.07}
\definecolor{darkgreen}{rgb}{0.0,0.6,0.00}

\begin{document}


\title{Local outflows from turbulent accretion disks}
\author{S\'ebastien Fromang \inst{1,2}, Henrik Latter \inst{3},
  Geoffroy Lesur \inst{4} and Gordon I. Ogilvie \inst{3}} 

\offprints{S.Fromang}

\institute{CEA, Irfu, SAp, Centre de Saclay, F-91191 Gif-sur-Yvette,
France \and UMR AIM, CEA-CNRS-Univ. Paris VII, Centre de Saclay,
F-91191 Gif-sur-Yvette, France \and Department of Applied Mathematics
and Theoretical Physics, University of Cambridge, Centre for
Mathematical Sciences, Wilberforce Road, Cambridge, CB3 0WA, UK \and
UJF-Grenoble 1 / CNRS-INSU, Institut de Planétologie et
d'Astrophysique de Grenoble (IPAG) UMR 5274, Grenoble, F-38041,
France. \\ \email{sebastien.fromang@cea.fr}}

\date{Accepted; Received; in original form;}

\label{firstpage}

\abstract
{}
{The aim of this paper is to investigate the properties of accretion
disks threaded by a weak vertical magnetic field, with a particular focus on
the interplay between MHD turbulence driven by the magnetorotational
instability (MRI) and outflows that might be launched from the disk.
}
{For that purpose, we use a set of numerical simulations performed
with the MHD code RAMSES in the framework of the shearing box
model. We concentrate on the case of a rather weak vertical magnetic
field such that the initial ratio $\beta_0$ of the thermal and magnetic
pressures in the disk midplane equals $10^4$.}
{As reported recently, we find that MHD turbulence drives an efficient
outflow out of the computational box. We demonstrate a strong sensitivity
of that result to the box size: enlargements in the radial and vertical
directions lead to a reduction of up to an order of magnitude in the
mass-loss rate. Such a dependence prevents any realistic estimates of
disk mass-loss rates being derived using shearing-box simulations. We
find however that the flow morphology is robust and independent of the
numerical details of the simulations. Its properties display some
features and approximate invariants that are reminiscent of the Blandford \& Payne
launching mechanism, but differences exist. For the magnetic field
strength considered in this paper, we also find that angular
momentum transport is most likely dominated by MHD turbulence, the
saturation of which scales with the magnetic Prandtl number, the ratio
of viscosity and resistivity, in a way that is in good agreement
with expectations based on unstratified simulations.}  
{This paper thus demonstrates for the first time that accretion disks
can simultaneously exhibit MRI--driven MHD turbulence along with
magneto-centrifugally accelerated outflows. However, in
contradiction with previously published results, such outflows
probably have little impact on the disk dynamics.}

\authorrunning{S.Fromang et al.}
\titlerunning{Stratified shearing boxes simulations of the MRI with vertical net flux}
\maketitle

\section{Introduction}
\label{introduction}

In the last two decades, MHD turbulence mediated by the
magnetorotational instability
\citep[MRI;][]{balbus&hawley91,balbus&hawley98} has been extensively
studied as the most likely mechanism responsible for angular momentum
transport in accretion disks. Yet the natural and simple
configuration of an initially uniform vertical magnetic field threading an isothermal 
vertically stratified disk has been relatively neglected compared to
the large number of papers that have been published reporting studies
of other configurations, such as unstratified disks, or stratified
disks with a toroidal magnetic field or without any net flux
\citep{stoneetal96,brandenburgetal95,miller&stone00,ziegler&rudiger00,fleming&stone03,hiroseetal06,johansen&levin08,oishi&mclow09,shietal10,davisetal10,flaigetal10,simonetal11a,simonetal11b,flaigetal12}. This
is largely the result of computational 
difficulties. Indeed, early simulations performed in the 1990s by
\citet{stoneetal96} and \citet{miller&stone00} found that the
configuration of a net vertical magnetic field in a stratified disk resulted in either the breakdown of their numerical codes or
the catastrophic disruption of the disk in a few dynamical times. These
disturbing findings, later confirmed by \citet{turneretal06}, led
researchers to focus on other, more user-friendly, configurations. This is
unfortunate because a configuration with a net poloidal magnetic field is the most
natural to explain the widely observed occurrence of jets and winds in 
astrophysical systems \citep[][and references
therein]{spruit96,wardle97}. Magnetic fields are also believed to 
be of key importance during the star-formation process
\citep{hennebelle&fromang08,hennebelle&ciardi09,commerconetal10,hennebelleetal11,seifriedetal11,joosetal12}. Even
if diffusion is required to remove much of the magnetic 
flux during that process \citep{mellon&li09,krasnopoletal10,krasnopoletal11,lietal11,santosetal12,dappetal12}, it is plausible that some net
poloidal flux is retained and carried down to the scale of the disk 
during its formation. The resulting strength of that potential magnetic
field is highly uncertain, but might be in the approximate range
$10^{-2}$ to a few Gauss \citep{wardle07}. This translates into $\beta$
values (the ratio between thermal and magnetic pressure) ranging from
$10^4$ to $1$.

Recently, however, the net vertical field configuration has received
renewed interest. Using local ideal MHD simulations performed in
the shearing-box approximation in that situation, \citet[][hereafter 
SI09]{suzuki&inutsuka09} found that accretion disks can drive winds
able to remove significant amounts of mass within dynamical
timescales. By varying the initial value of the plasma $\beta$
parameter, the ratio of midplane thermal 
and magnetic pressures, between $10^4$ and infinity (i.e. no vertical
magnetic field), they found that the mass outflow rate driven by such
winds is 
an increasing function of the magnetic field strength. For the
strongest field they investigated, $\beta=10^4$, the outflow is so
powerful that the disk is emptied in less than $100$ local
orbits! As investigated later by the same authors
\citep[][hereafter SMI10]{suzukietal10}, such strong disk outflows have important
consequences for disk evaporation and could also help halt inward
planet migration. This is an important result that, as such, 
requires confirmation and follow-up. This is the first goal of this
paper.

Another important unresolved issue under these conditions is the 
saturation of the turbulence itself. Using unstratified shearing
boxes, early simulations \citep{hawleyetal95} soon established that
the rate of angular momentum transport scales like the initial
vertical magnetic field strength. Note however that this scaling has
recently been questioned by various authors
\citep{bodoetal11,uzdensky12}. In addition, extensive parameter surveys 
\citep{lesur&longaretti07,longaretti&lesur10} performed using the same
setup also demonstrated a strong dependence of the transport with the magnetic
Prandtl number $Pm$, the ratio of kinematic viscosity to ohmic
resistivity. How these results are modified when taking vertical
stratification into account is still unknown. Performing a first
step in that direction is the second goal of that paper. As recently
argued by \citet{uzdensky12}, the question of the saturation of MHD
turbulence in 
such a situation is not only important for angular momentum transport
itself, but is also linked to issues such as disk coronal heating and
magnetic flux transport.

The plan of the paper is as follows. In
section~\ref{numerical_method_sec}, we present the numerical method we 
used and describe the extensions of the basic scheme that were required
for the simulations to run without failure. The immediate result of these
simulations is a confirmation of the results reported by SI09: an MHD
turbulent disk threaded by a net vertical flux drives outflows that
efficiently remove mass from the computational box. In
section~\ref{num_issues_sec} we focus on the sensitivity of the
results to details of the numerical setup, such as resolution and box
size. Finally, in section~\ref{flow_prop_sec}, we make a connection with
existing disk wind theories and study the saturation properties of the
turbulence, before discussing some aspects of our results 
(section~\ref{discussion_sec}) and perspectives for future work
(section~\ref{conclusion_sec}).

\section{Methods}

\subsection{Numerical setup}
\label{numerical_method_sec}


\noindent
The numerical simulations we present in this paper use
a setup similar to that of \citet{stoneetal96} and
\citet{brandenburgetal95}. The MHD equations are solved in a Cartesian
coordinate system $(x,y,z)$ with unit vectors $(\bb{i},\bb{j},\bb{k})$
rotating with angular velocity $\Omega$ around a central mass. This is
the classical shearing-box model
\citep{goldreich&lyndenbell65,hawleyetal95}. As described by previous
authors, these equations include a vertical component of
the gravitational force and possibly also viscous and ohmic
dissipation \citep[see also][]{davisetal10}. For simplicity, we
consider only the case of an isothermal gas with sound speed
$c_0$. This is an important limitation as it precludes the
possibility of coronal heating in the disk atmosphere, which is
potentially important for the question of outflow launching. It also
artificially increases the role of thermal pressure in launching
possible outflows by preventing cooling of accelerating gas. More 
realistic vertical thermodynamic structure of the disk should be
considered in future work.

As described in the introduction, the aim of the present paper is to
examine configurations in which the vertical magnetic flux (conserved
during the simulation because of the shearing box symmetries) does not
vanish. This is done by initializing the magnetic field as a uniform
vertical field $\bb{B}=B_0 \bb{k}$ whose strength is parametrized by
the midplane plasma parameter $\beta$ defined according to
\begin{equation}
\beta_0=\frac{\rho_0 c_0^2}{B_0^2/2}
\end{equation}
where $\rho_0$ is the value of the gas density $\rho$ in the disk
midplane, and we work in electromagnetic units such that
$\mu_0=1$. In this paper, we consider the case $\beta_0=10^4$. 
Such a value corresponds to the strongest field considered by
\citet{suzuki&inutsuka09} and leads to large mass outflow rates as
discussed in the introduction. Recently, \citet{lesuretal12},
\citet{bai&stone13} and \citet{moll12} considered lower values of
$\beta_0$, thus providing a complete coverage of the behaviour of the
flow as a function of magnetic field strength up to equipartition.
The purely vertical magnetic field we start with here is superposed on
the initial hydrostatic disk configuration at the beginning of the
simulations, along with velocity fluctuations that trigger the growth
of the MRI. For such a value of $\beta_0$, gas and magnetic field
are in equipartition at $z \sim 4.3 H$, where $H=c_0/\Omega$ defines
the disk scaleheight. 

We used standard shearing-box boundary conditions in the radial
direction \citep{hawleyetal95} and periodic conditions in the
azimuthal direction. At the vertical boundaries, however, we used
modified outflow boundary conditions. The density is extrapolated
assuming vertical hydrostatic equilibrium as described in
the appendix~A.2 of \citet{simonetal11a}. Zero-gradient boundary
conditions are applied on horizontal velocities and on the vertical
momentum, when matter is outflowing (otherwise, the vertical
velocity is set to zero in the ghost cells). Finally, the magnetic
field is forced to be vertical in the ghost cells.

The set of equations just described are solved using a version of
the code RAMSES \citep{teyssier02,fromangetal06} that solves the MHD
equations on a uniform grid in the shearing box. It was
quantitatively tested by \citet{latteretal10} and the results obtained
in 3D simulations of MRI--driven turbulence were shown to 
compare successfully with those obtained with the code Athena \citep{stoneetal08}
by \citet{fromang&stone09}. In the present paper, several simulations
are performed in a radially extended box (larger than the disk scale
height $H$). \citet{johnsonetal08} have shown that position-dependent 
dissipation introduces artificial radial variations of the density and
the stress in this case. As shown by the same authors, this effect can
be eliminated by using an MHD extension of the FARGO orbital advection
algorithm \citep{masset00}. For the purpose of this work, we have
implement such an extension, following the method recently described
by \citet{stone&gardiner10} for finite-volume methods.  

\subsection{Extensions of the basic scheme}

\noindent
Shearing-box numerical simulations that start with vertical magnetic
flux lead to vigorous turbulence. Strong magnetic fields develop in
the upper layers of the disk and can cause the code to crash randomly
during a run. For this reason, a number of modifications had to be
implemented for the simulations to proceed robustly. We
describe them here. 

First, RAMSES uses the MUSCL-Hancock algorithm to integrate the MHD
equations \citep{toro97}. Spatial slopes of all variables are required
for the scheme to be of second order in space and time. Simulations
performed with RAMSES normally use the MinMod or MonCen 
slope limiters \citep{toro97}. However, we found that both limiters
systematically produce evacuated regions where the Alfv\'en velocity is
enormous, resulting in extremely small time steps that cause the code
to halt. We therefore used the multidimensional limiter described by
\citet{suresh00} that is able to avoid such problems.

However, this turned out to be insufficient in radially extended
boxes. Two additional modifications were implemented. First, at
those cell interfaces (usually located in the disk corona) where the
magnetic pressure exceeds the thermal pressure by more than three
orders of magnitude, we used the Lax--Friedrichs Riemann solver. The HLLD
Riemann solver \citep{miyoshi&kusano05} is used at all other
locations. In addition, a mass diffusion source term was added to the
continuity equation. While conserving mass, it helps fill the density
holes described above and prevent large Alfv\'en speeds from appearing. We
used the same analytical expression for the diffusion coefficient as
\citet{gresseletal11}:
\begin{equation}
D=D_0 \left[ 1 + \left( \frac{\rho}{\rho_0} 10^{C_{dyn}}\right)^4 \right]^{-1/4},
\end{equation}
with $D_0=10^{-2} c_0^2/\Omega$ and $C_{dyn}=5$. Such parameters yield
a grid Reynolds number for mass diffusion of order unity in the disk
upper layers. 

When used simultaneously, the modifications to the standard scheme
described in this section permit a robust execution of the code until
completion of the simulations. The modifications they introduce to the
flow are undetectable.

\subsection{Run properties}

\begin{table*}[t]\begin{center}\begin{tabular}{@{}ccccccccc}\hline\hline
Model & Box size & Resolution & Run time (orbits) & Re & Rm & $\dot{m}_w$ & $\alpha_{SS}$ \\
\hline\hline
ZRm3000 & $(4H,4H,H)$ & $(256,128,64)$ & 100 & 3000 & 3000 & - & $4.0 \times 10^{-2}$ \\
ZRm1500 & $(4H,4H,H)$ & $(256,128,64)$ & 100 & 3000 & 1500 & - & $2.4 \times 10^{-2}$ \\
\hline
SI09run  & $(\sqrt{2}H,4\sqrt{2}H,8\sqrt{2}H)$ & $(32,64,256)$ & 20 & - & - & $2.4 \times 10^{-2}$ & $8.5 \times 10^{-2}$ \\ 
\hline
Ideal256  & $(H,4H,10H)$ & $(32,64,256)$ & 20 & - & - & $2.6 \times 10^{-3}$ & $1.0 \times 10^{-1}$ \\ 
Ideal512  & $(H,4H,10H)$ & $(32,64,512)$ & 20 & - & - & $2.7 \times 10^{-3}$ & $1.4 \times 10^{-1}$ \\ 
Ideal1024 & $(H,4H,10H)$ & $(32,64,1024)$ & 20 & - & - & $2.3 \times 10^{-3}$ & $1.1 \times 10^{-1}$ \\ 
Ideal2048 & $(H,4H,10H)$ & $(32,64,2048)$ & 20 & - & - & $2.8 \times 10^{-3}$ & $1.2 \times 10^{-1}$ \\ 
\hline
Diffu1H & $(H,4H,10H)$ & $(64,128,640)$ & 30 & 3000 & 3000 & $2.2 \times 10^{-3}$ & $8.3 \times 10^{-2}$ \\ 
Diffu2H & $(2H,4H,10H)$ & $(128,128,640)$ & 30 & 3000 & 3000 & $1.5 \times 10^{-3}$ & $7.5 \times 10^{-2}$ \\ 
Diffu4H & $(4H,4H,10H)$ & $(256,128,640)$ & 40 & 3000 & 3000 & $1.1 \times 10^{-3}$ & $6.2 \times 10^{-1}$ \\ 
Diffu8H & $(8H,8H,10H)$ & $(512,256,640)$ & 10 & 3000 & 3000 & $1.1 \times 10^{-3}$ & - \\ 
\hline
Tall4H & $(4H,4H,20H)$ & $(256,128,1280)$ & 20 & 3000 & 3000 & $3.2 \times 10^{-4}$ & $7.8 \times 10^{-2}$ \\ 
\hline
Diffu4Ha & $(4H,4H,10H)$ & $(256,128,640)$ & 40 & 3000 & 1500 & $1.2 \times 10^{-3}$ & $4.8 \times 10^{-2}$ \\ 
\hline\hline
\end{tabular}
\caption{Properties of the runs described in this paper. The first column
gives the model label, while the next two columns give the size of the
computational domain $(L_x,L_y,L_z)$ and the resolution
$(N_x,N_y,N_z)$ of that run. We give in column four the
duration (in orbits) of each simulation. Dissipation coefficients,
expressed in terms of the 
Reynolds and magnetic Reynolds numbers, are given in the following two
columns. The following column gives, in dimensionless form, the
rate $\dot{m}_w$ at which 
matters escapes the computational domain per unit surface
area. Finally, the last column provides the time--averaged value of
$\alpha_{SS}$ for all models.}
\label{runProperties_tab}
\end{center}
\end{table*}

\noindent
We summarize in Table~\ref{runProperties_tab} the properties of
the simulations presented in this paper. Table~\ref{runProperties_tab}
shows the model label (column 1), the box size (column 2), the grid
resolution (column 3) and the run duration (column 4). Columns 5 and 6
report, where applicable, the Reynolds and magnetic Reynolds numbers we
used. They are respectively defined by:
\begin{equation*}
Re=\frac{c_0 H}{\nu} \textrm{ and } Rm=\frac{c_0 H}{\eta} \, ,
\end{equation*}
where $H=c_0/\Omega$ is the disk scale height while $\nu$ and $\eta$
are the kinematic viscosity and ohmic resistivity (magnetic diffusivity)
respectively. Finally, the average dimensionless mass-loss rate per
unit area through the 
upper and lower surfaces of the box $\dot{m}_w$ is reported in
column 7. It is calculated according to
\begin{equation}
\dot{m}_w=\frac{1}{2}\frac{(\rho v_z)_{top}+(\rho v_z)_{bot}}{\rho_0
  c_0 L_x L_y} \, ,
\label{mdot_w_eq}
\end{equation}
where $(\rho v_z)_{top}$ and $(\rho v_z)_{bot}$ are respectively the
horizontally integrated mass fluxes through the lower and upper
surfaces of the box. The rate of angular momentum transport,
expressed in term of the parameter $\alpha_{SS}$, is given in
the last column for all models. It is calculated as the sum of the
height--integrated Reynolds and Maxwell stresses, normalized by the
height--integrated pressure:
\begin{equation}
\alpha_{SS}=\frac{\int \mean{\rho \delta v_R \delta v_{\phi}} - \mean{B_R
    B_{\phi}} dz }{\int \mean{\rho} c_0^2 dz } \, ,
\end{equation}
where $\mean{.}$ denotes an horizontal average.

\section{Numerical issues}
\label{num_issues_sec}

In all the simulations we performed using the setup described above,
we found that strong disk outflows similar to those described by
SI09 and SMI10 develop. This indicates that outflows are a robust
outcomes of MHD shearing-box simulations of stratified disks threaded
by a vertical magnetic field. In this section we analyse in detail the
influence of various computational parameters, such as vertical
boundary conditions, resolution and box size, on the properties of
these outflows.

\subsection{Comparison with \citet{suzuki&inutsuka09}}

As described in section~\ref{numerical_method_sec}, we use vertical
boundary conditions that are different from SI09. Indeed, these
authors prescribed outflow conditions by adopting only outgoing MHD
characteristics at their top and bottom boundaries
\citep{suzuki&inutsuka06}. Our choice of different 
boundary conditions provides an opportunity to test the sensitivity
of the outflows to this aspect of the numerical setup. To assess this
possibility in a quantitative way, we 
reproduce their model as closely as possible by performing a simulation in a box of
the same size $(L_x,L_y,L_z)=(\sqrt{2}H,4\sqrt{2}H,8\sqrt{2}H)$ and
with identical resolution $(N_x,N_y,N_z)=(32,64,256)$. The $\sqrt{2}$
factor in the box size appears because of our different definition of
$H$ compared to SI09 (we will later return to box sizes that are
integer numbers of $H$). This model is labelled {\it SI09run} (see 
Table~\ref{runProperties_tab}). Note that the simulation was only
evolved for about $20$ orbits, which is short by modern standards.  However,  $20 \%$ of the disk mass has escaped the
computational box by that time, and we found that this is large enough to
affect the subsequent flow properties. The mass-loss rate per unit
area $\dot{m}_w$ we measured in this model, averaged in time between
$5$ and $20$ orbits, is $2.4 \times 10^{-3}$. SI09 do not
quote explicitly their values, which can however be read off from their
figures. In their figure 5, SI09 report $2 \dot{m}_w \sim 5$--$6 \times
10^{-3}$. Thus there is good agreement with our results, which
suggests  the particular choice of boundary conditions has only a
limited effect on the quantitative properties of the flow. 


\subsection{Influence of the vertical resolution}
\label{resolution_sec}

\begin{figure}
\begin{center}
\includegraphics[scale=0.45]{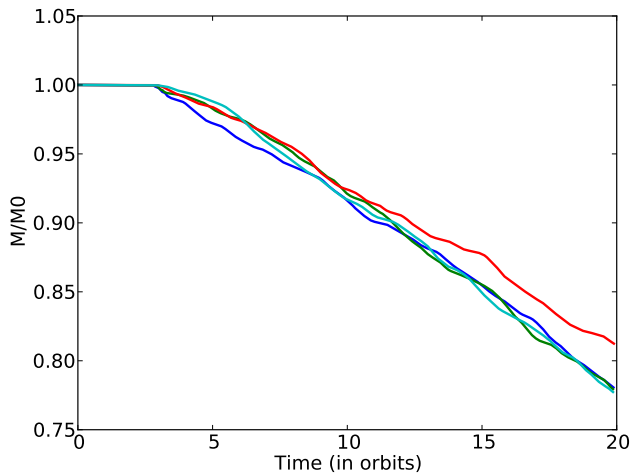}
\caption{Time evolution of the total relative mass in the box when the
number of cells in the vertical direction is varied: $N_z=256$ ({\it blue line}),
$N_z=512$ ({\it green line}), $N_z=1024$ ({\it red line}) and
$N_z=2048$ ({\it light blue line}). All other run parameters are fixed.}
\label{massVsTimeJaps_fig}
\end{center}
\end{figure}

A peculiar aspect of SI09 is the low
resolution they used, as noted by \citet{guan&gammie11}. Indeed, the
most unstable MRI mode has a vertical wave number given by $k_z v_{Az}
\sim \Omega$ in an unstratified box. For $\beta=10^4$, as considered
in the present paper and by SI09, this
corresponds to a wavelength of about one tenth of the disk
scale height. This wavelength is not adequately resolved in model {\it ST09a} in which we use
$32$ cells per $H$. In stratified disks, the eigenmode structure is more
complex and cannot be reduced to a single wave number, but
\citet{latteretal10} have shown that about $200$ cells per scale height
are required in this case to properly resolve the fastest growing
eigenmode across the whole disk. In order to assess the importance of
the vertical resolution on the existence and strength of the outflows
that develop in model {\it ST09a}, we computed four models ({\it
  Ideal256} to {\it Ideal2048}), gradually increasing the vertical 
resolution 
from $N_z=256$ to $N_z=2048$ while keeping all other parameters
fixed. In figure~\ref{massVsTimeJaps_fig}, we plot the time-evolution
of the total gas mass $M$ in the computational box, normalized by its
initial value $M_0$. We find it is similar in the four models,
regardless of the number of vertical cells. The values of $\dot{m}_w$
for the four models are reported in
Table~\ref{runProperties_tab}. They confirm the results of
figure~\ref{massVsTimeJaps_fig}, as all values agree within $\pm
15\%$, with most of this scatter being due to model Ideal1024
for which we obtained a smaller mass loss rate than the other
models. As a matter 
of fact, there is no systematic trend in the evolution of $\dot{m}_w$
with resolution. We conclude that there is only a modest effect of
resolution, if any, on the result. It is tempting to associate
this weak dependence with the
morphology of the eigenmodes coupled with the fact that disk outflows
are launched at $z\sim \pm 2 H$, as SI09 and SMI10 do. Indeed,
MRI modes have a shorter wavelength near the midplane and a longer
wavelength at greater height \citep{latteretal10}. Since outflow rates
are not set by conditions at the midplane, the mode structure at the
midplane may be underresolved with little impact. Extending
the classical result of \citet{goodman&xu94} to stratified boxes, 
\citet{latteretal10} have also shown that those modes are not only
solution of the linear equation, but also nonlinear solutions in the
limit of large $\beta$, such that some of their properties might
affect the nonlinear flow dynamics. This reasoning would also agree
with the claim of SI09 that such disk outflows are due to the breakup
of channel modes in the disk corona. It is indeed likely that 
such channel modes will help the build up of strong fields in
the disk atmosphere as is observed during the linear stage of the
MRI. That having being said, the dynamics of the disk atmosphere
during the remainder of the simulations is highly nonlinear and
fluctuating. It might share some properties of the dynamics of channel
modes, but a direct relationship between such outflows and linear MRI
modes remains difficult to assess with certainty.

\subsection{Explicit dissipation coefficients}

The previous section has established that the mass outflow rate
present a modest but apparently non--monotonic dependence with grid
resolution. In addition, such simulations rely on grid resolution in
order to stabilize the numerical scheme, the form and effects
of which are more difficult to quantify than those of explicit
dissipation. As explained in the introduction, it is also well known
that the saturation properties of the turbulence depend strongly on
the dissipation coefficients
\citep{lesur&longaretti07,longaretti&lesur10,fromangetal07,fromang10,simon&hawley09,davisetal10,simonetal11a},
regardless of the magnetic field configuration. 
 
All these reasons argue in favor of including explicit dissipation in such
simulations. Their introduction also makes the problem mathematically
well posed. In contrast to the low-vertical-resolution simulations
described above (for which numerical dissipation influences the linear
phase of the instability), when viscosity and Ohmic resistivity are
included the calculated growth rates and morphology of all the active
MRI normal modes can be accurately reproduced in the simulations,
providing a valuable code test. In the appendix, we describe the
calculation of these modes as a function of $Re$ and $Rm$, extending
the method used by \citet{latteretal10} in the ideal MHD limit. As shown in
fig.~\ref{eigenmode_fig}, the linear modes display large-scale
oscillations at $z \sim \pm 3H$ regardless of $Re$ and $Rm$, while the
small-scale oscillations near the midplane are very much influenced 
by the magnitude of the dissipation coefficients. This
structure could also explain the relative independence of the flow
structure on resolution described in section~\ref{resolution_sec}.

One obvious limitation associated with explicit dissipation
coefficients, though, is the tremendous computational costs associated
with the grid resolution that is required for the dissipative
lengthscales to be properly resolved. To keep the resolution
requirements reasonable, we 
adopt in the remainder of this paper rather modest values for the
dissipation coefficients such that $Re=3000$ for all simulations. We
consider two values for $Rm$, namely $Rm=3000$ in our fiducial model
and $Rm=1500$ in section~\ref{turb_transport_sec}. Past experience
gained from unstratified simulations \citep[see for
example][]{fromangetal07} suggests that $64$ cells per disk scale height
are sufficient to capture correctly the flow properties at all
relevant scales. The vertical profile of the eigenmode in such a
case is illustrated in the bottom panel of
fig.~\ref{eigenmode_fig}. Small scale oscillations near the midplane
are damped by the large viscosity and resistivity. As a result, the
most unstable eigenmode is easily captured even with such a modest
resolution. This is another benefit gained from introducing
dissipation in the problem. Before moving to a systematic study of the
flow in such a 
viscous and resistive model, we briefly compare the outflow rate
obtained in one such case (which we label model {\it Diffu1H} for future
reference) with the cases without explicit dissipation described above, all other parameters being
unchanged. As shown in Table~\ref{runProperties_tab}, $\dot{m}_w$ is
similar in all cases, regardless of the nature of the
dissipation. This is in agreement with the above result that mass 
outflow rates are independent of the resolution, when explicit
dissipation is absent. We will therefore 
adopt the setup of model {\it Diffu1H} as our fiducial model in the
following and we now move to an extensive study of the disk outflow 
properties and saturation level of the turbulence using these
parameters.  

\subsection{Influence of box size}

\subsubsection{Radial box size}
\label{Lx_sec}

\begin{figure}
\begin{center}
\includegraphics[scale=0.45]{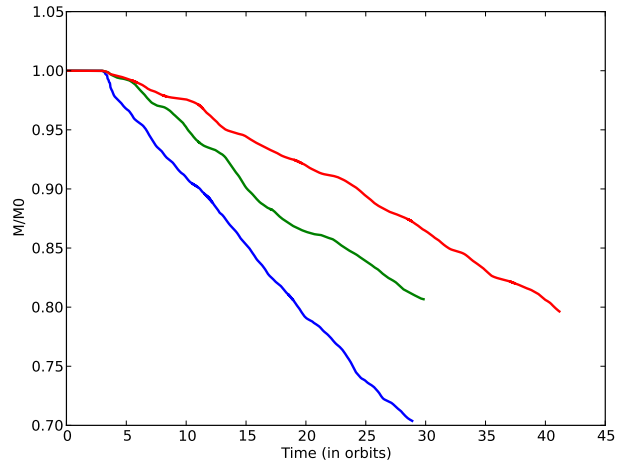}
\caption{Time-evolution of the total relative mass in the box when the
radial extent of the box is varied from $L_x=H$ ({\it blue line}) to
$L_x=2H$ ({\it green line}) and $L_x=4H$ ({\it red line}).}
\label{massVsTimeLx_fig}
\end{center}
\end{figure}

\begin{figure}
\begin{center}
\includegraphics[scale=0.45]{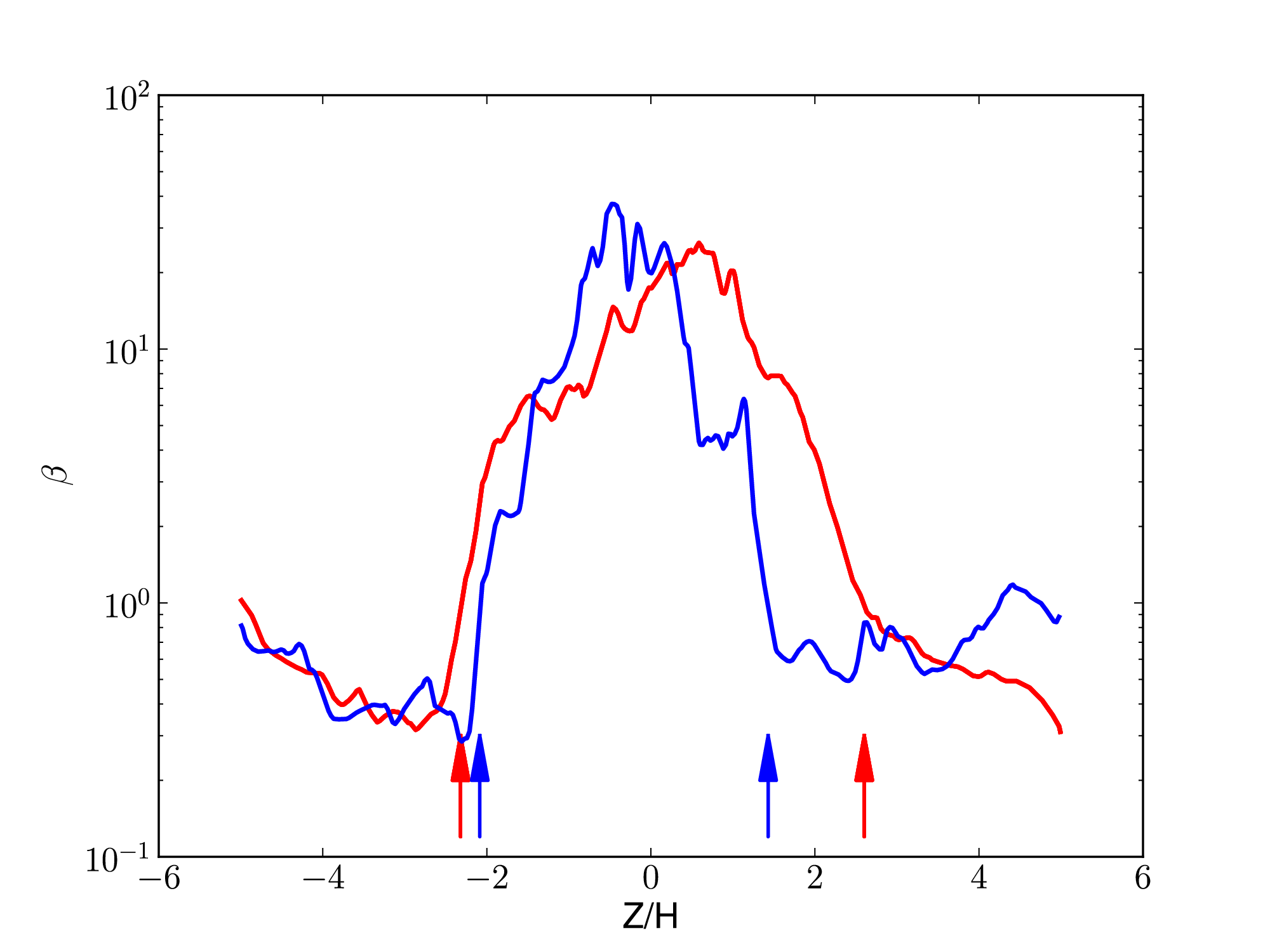}
\caption{Vertical profile of the ratio $\beta$ between the horizontally
  averaged thermal pressure and the horizontally averaged magnetic
  pressure. The blue curve corresponds to model {\it Diffu1H} and
  results from a further time average between $t=10$ and $t=15$
  orbits. The red curve was obtained averaging the data of model {\it
    Diffu4H} between $t=10$ and $t=30$. For each model, the vertical
  arrows report the approximate wind launching location where $\beta=1$
  (SI09).}
\label{betaVsZ_fig}
\end{center}
\end{figure}


\noindent
In a suite of numerical simulations of unstratified shearing boxes
threaded by a net vertical flux, \citet{bodoetal08} discovered that
the radial extension of the domain can significantly affect the
properties of the turbulence. They demonstrated that channel flows
develop in boxes of unit radial extent ($L_x=H$) while their influence is
reduced in boxes having larger $L_x$. Since SI09
argue that {\it the breakup of these channel flows drives structured disk
winds}, it is natural to ask whether the radial extent of the shearing
box affects the disk outflow properties described above. This is the
purpose of the present section.

To do so, we performed a series of simulations,
successively doubling the radial extent of the box from $L_x=1H$ (model
{\it Diffu1H}) to $L_x=4H$ (model {\it Diffu4H}). For each model, the
time-evolution of the total mass inside the computational domain is shown
in figure~\ref{massVsTimeLx_fig}. The mass-loss rate gradually
decreases as $L_x$ is increased. This is confirmed by
Table~\ref{runProperties_tab}, where $\dot{m}_w$ is found to be reduced by a factor of $2$ in the largest model. Such a lower mass-loss rate, apart
from changing the properties of the outflow itself, also allows the
simulation to be run for longer (and more statistics of the
turbulence to be accumulated) before the reduction of the disk surface
density starts affecting the disk structure. This is the reason why we
could run model {\it Diffu4H} for $40$ orbits. A model with $L_x=8H$
(model {\it Diffu8H}) was also run and suggested numerical 
convergence for larger radial extent. However, its large resolution,
$(N_x,N_y,N_z)=(512,256,640)$, resulted in huge computational
requirements that prevented the simulation from being run for more than 10
orbits. Clearly, this question needs further investigation but our preliminary
result is not in disagreement with the conclusions of
\citet{bodoetal08}, who find good agreement between their $L_x=4$ and
$L_x=8$ runs.

For a better understanding of the differences between these
simulations, we turn to an examination of the wind launching
location. SI09 have shown that this position can be
evaluated to a good approximation by computing the location where the
ratio $\beta$ between the horizontally averaged thermal pressure and
the horizontally 
averaged magnetic pressure falls below unity. The vertical profile of
$\beta$ is plotted in figure~\ref{betaVsZ_fig} for models {\it
  Diffu1H} and {\it Diffu4H}. For both models, $\beta$ is large in the
turbulent midplane and decreases upwards. The location $z_{w}$
where it falls below unity is reported using vertical arrows. For the
smallest radial box size, we measured $z_w=-2.1H$ and $2.3H$
respectively below and above the equatorial plane. The same values are
$z_w=-2.7H$ and $2.9H$ when $L_x=4H$. Thus, in bigger boxes, disk
outflows are launched at higher altitude, where the gas density is
smaller, resulting in a smaller mass-loss rates. Ultimately, the
higher location of the launching region can be be traced back to the
toroidal magnetic field being weaker in the disk upper layers ($z \sim
2-3 H$) in bigger boxes by a factor of about two, while the gas density
varies by only a few percent at those locations. The physical
mechanism responsible for this reduction of the toroidal magnetic
field is uncertain, but could be due to the weakening of coherent
structures as demonstrated by \citet{bodoetal08} in unstratified
boxes. This weakening indeed results in more disordered magnetic
fields (seen in the simulation as a larger ratio between the
fluctuating and means parts of the azimuthal magnetic field) that
reduce the efficiency with which outflows are accelerated. 

\subsubsection{Vertical box size}
\label{Lz_sec}

\begin{figure}
\begin{center}
\includegraphics[scale=0.45]{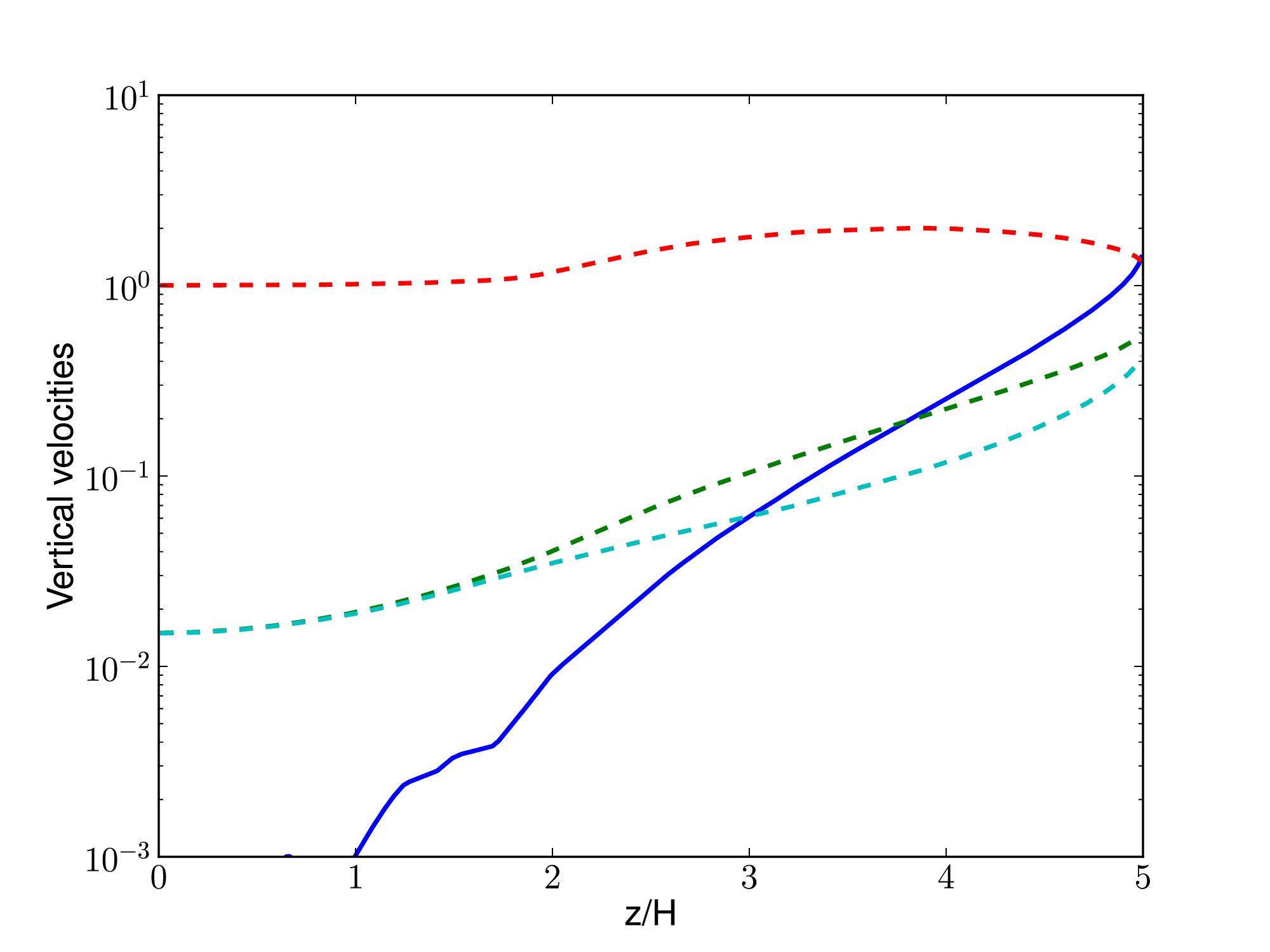}
\includegraphics[scale=0.45]{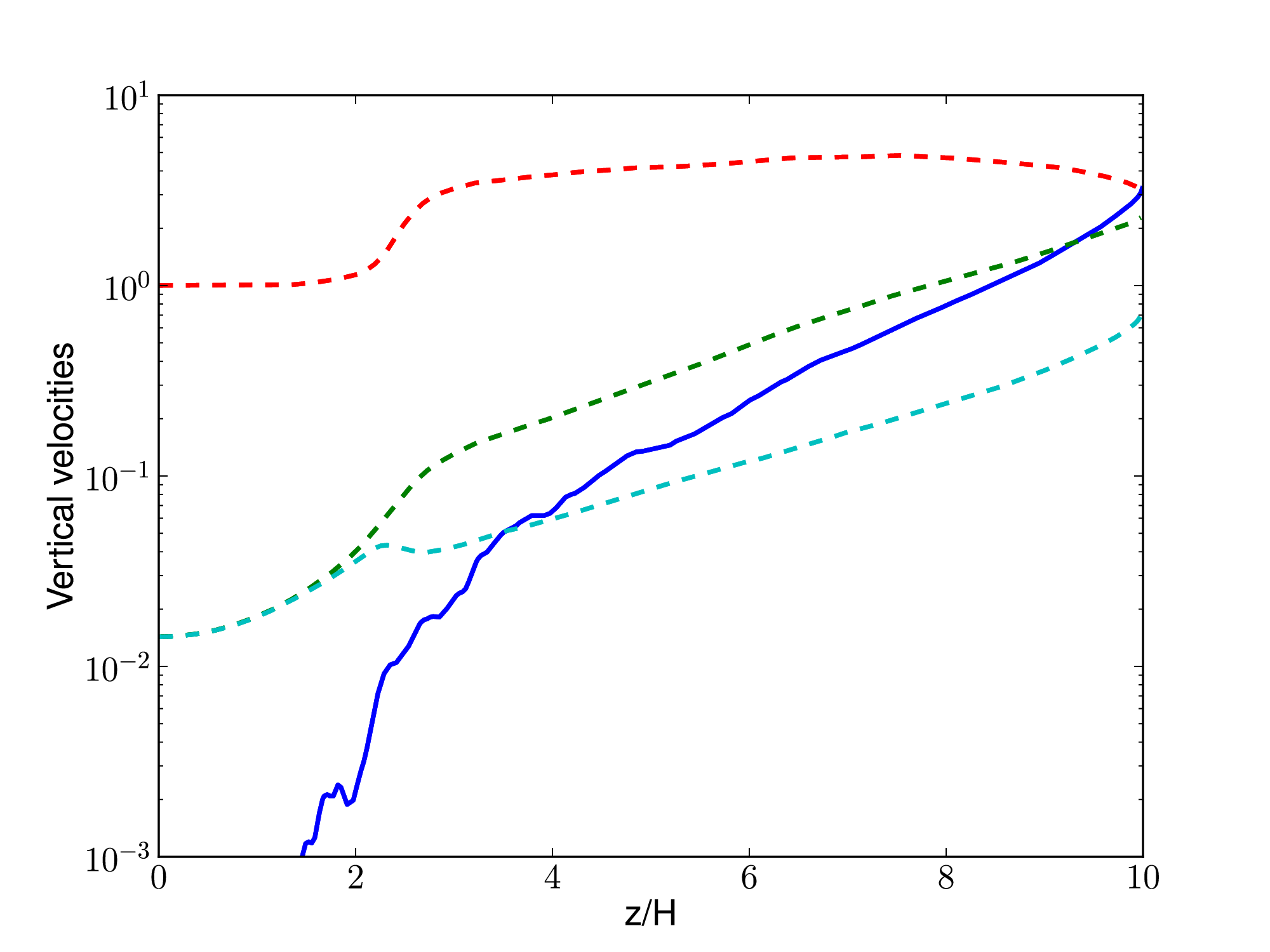}
\caption{Vertical profiles of the vertical fluid and wave velocities.  Horizontally and time-averaged fluid velocity $v_z$
({\it blue curve}), slow magnetosonic speed $v_{sz}$ ({\it light blue dashed curve}), Alfv\'en speed $v_{Az}$
({\it green dashed curve}) and fast magnetosonic speed $v_{fz}$ ({\it red dashed curve}), in units of $c_0$. The
top panel shows results obtained in model {\it Diffu4H}
($L_x=4H$, $L_z=10H$) while the bottom panel is for the model {\it Tall4H}
($L_x=4H$, $L_z=20H$).} 
\label{sonicPoints_fig}
\end{center}
\end{figure}

\noindent
In numerical simulations, it is desirable for the outflow velocity
to exceed all wave velocities before leaving the computational
domain. When this occurs, no signal can propagate downwind, which
guarantees that the results are independent of the box size and the
boundary conditions.\footnote{We note, however, that such a situation 
does not guarantee a successful launching of the outflow, as this
depends on its ability to escape from the gravitational potential. We
will come back to this point in section~\ref{wind_transport_sec}.} In
the case of the magnetized outflows 
studied here, the relevant wave velocities are those of the Alfv\'en
wave and the slow and fast magnetosonic waves, each of which gives
rise to a critical point in the flow \citep{spruit96}. Given the
complexity of the flow, we will not consider wave velocities parallel
to the fluid streamlines in this paper, but rather will focus on
vertical velocity and wave velocities relative to the fluid. In the
published simulations of SI09 and SMI10, no information 
is given as to whether the critical points are crossed, but this is
probably not the case, because the flow velocity remains subsonic in
some of the cases they report (see for example figure 3 of SI09). 

In the present section, we investigate this issue by comparing the
results of model {\it Diffu4H}, described above, with those of model
{\it Tall4H}. The latter differs from the former only in its vertical
extent, namely $L_z=20H$. We emphasize that the resolution of that
simulation is large: $(N_x,N_y,N_z)=(256,128,1280)$, which amounts to
40 million cells. Thus the computational requirements are significant
and that model was evolved for only $20$ orbits. This is
however sufficient to derive a reliable estimate of the mass-loss rate, as
shown by both figures~\ref{massVsTimeJaps_fig} and
\ref{massVsTimeLx_fig} for the models we have described so far. We
found that the mass outflow is significantly reduced when 
doubling the vertical extent of the box. Indeed,
table~\ref{runProperties_tab} shows that $\dot{m}_w$ is reduced by a 
factor of $3.4$. This is not a consequence of the mass flux
steadily decreasing with $z$, as we observe in all simulations a
constant $\mean{\rho v_z}$ above $z \sim 3H$, as reported also by
\citet{bai&stone13}. On the other hand, the existence of vertical
boundaries could be preventing inflowing velocities. Indeed,
both the 10H and 20H simulations display comparable outflowing mass
fluxes at $z \pm 5H$, when only accounting for gas moving away from
the midplane. This suggests that boxes with smaller
vertical extents could lose material that would otherwise fall back on
to the disk, thus artificially enhancing $\mean{\rho v_z}$. However
this effect is absent by construction in the $1D$ simulations
performed by \citet{lesuretal12} where the same dependence with
vertical box size is obtained, thus questionning such an
interpretation for the origin of the decrease.

Here, we consider an alternative explanation by investigating the
location of the outflow critical points. In 
figure~\ref{sonicPoints_fig}, the vertical profile of the  
vertical velocity (normalized by the sound speed) is compared with
the vertical profile of the vertically propagating wave velocities
(also normalized by $c_0$) for
models {\it Diffu4H} ({\it top panel}) and {\it Tall4H} ({\it bottom
  panel}). The latter are given by the solutions of the following
expression \citep{ogilvie12}:
\begin{equation}
\left[ v^2-v_{az}^2 \right] \left[ v^4-(c_0^2+v_a^2)v^2+c_0^2v_{az}^2 \right]=0 \, ,
\end{equation}
where $\bb{v_a}=\bb{B}/\sqrt\rho$ is the Alfven velocity.
Two comments can be made. In both panels, it is seen that only the
slow and Alfv\'en 
points are crossed, while the fast magnetosonic point is, suspiciously,
reached at the vertical boundary itself. The fact that this is 
the case in both models suggest that this is no coincidence, but is
rather a numerical artifact probably associated with the shearing-box
model or simply with the limited vertical extent of the
computational domain. We note that such a failure to cross the
fast magnetosonic point is not unlike published simulations of
magnetically driven jets \citep[][see their
fig. 1]{casse&ferreira00} and has also recently been reported by
\citet{lesuretal12} and \citet{bai&stone13} in simulations using a
similar setup but performed 
with a different code. A second remark is that the slow and Alfv\'en
points lie systematically higher up in the disk atmosphere in the
taller model. The non-convergence of the location of those points
might explain the different mass outflow rates measured in the two
simulations \citep{spruit96}.

We conclude this section by stressing the importance of this
effect. The sensitivity to box size we obtain means that while $20 \%$ of the  
mass in the box is lost within $20$ orbital times in the simulations of
SI09, it would take about $140$ orbits to lose
the same amount of mass in model {\it Tall4H}. The consequences for
disk evolution would surely be significant.


\section{Flow properties}
\label{flow_prop_sec}

As shown in both sections~\ref{Lx_sec} and \ref{Lz_sec}, it
appears that many details of the computational setup affect the
properties of outflows from MHD turbulent disks. The disappointing but
unavoidable conclusion from these results is that the disk mass-loss 
rate due to such outflows cannot be robustly inferred in the framework of
the shearing-box model. Nevertheless, we found that a number of
properties of the flow are displayed by all simulations. The purpose
of the present section is to describe these features and to analyse
quantitatively the angular momentum transport associated with the
turbulence.

However, one should be cautious when doing so. As already mentioned,
the strong mass 
loss due to the outflow changes the disk structure itself on
timescales of a few tens of orbits. For example, in model {\it
Diffu1H}, we discovered that a strong mean toroidal field, with a
strength close to equipartition, emerged in the disk
midplane after $30$ orbits, perturbing its structure thereafter. No such 
perturbation was evident in model {\it Diffu4H}. In 
fact, all the diagnostics we considered suggest the disk is in a
quasi-steady state during the first $20$ to $25$ orbits that follow
the establishment of fully developed turbulence. This is why we will
restrict our analysis in this section to the model having
$L_x=4H$. Unless otherwise specified, time averages of the data will
cover the range $5<t<30$.

\subsection{Disk atmosphere: outflow structure}
\label{outflow_struc_sec}

\begin{figure}
\begin{center}
\includegraphics[scale=0.43]{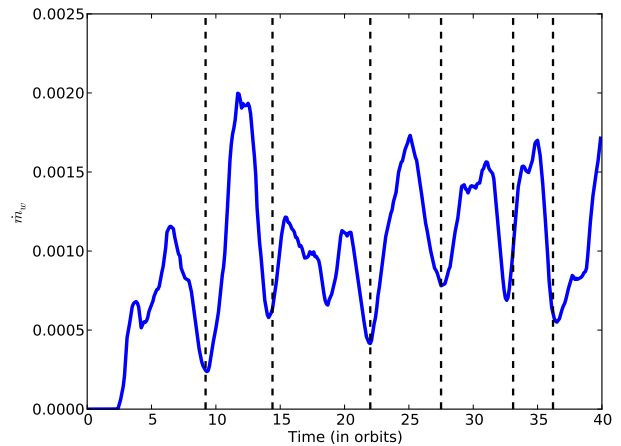}
\caption{Time evolution of $\dot{m}_w$ for
model Diffu4H ({\it blue curve}). The vertical lines mark the times at
which the mean toroidal magnetic field at the top boundary of the box
changes sign.} 
\label{massFluxLx4h_fig}
\end{center}
\end{figure}

\begin{figure*}
\begin{center}
\includegraphics[scale=0.43]{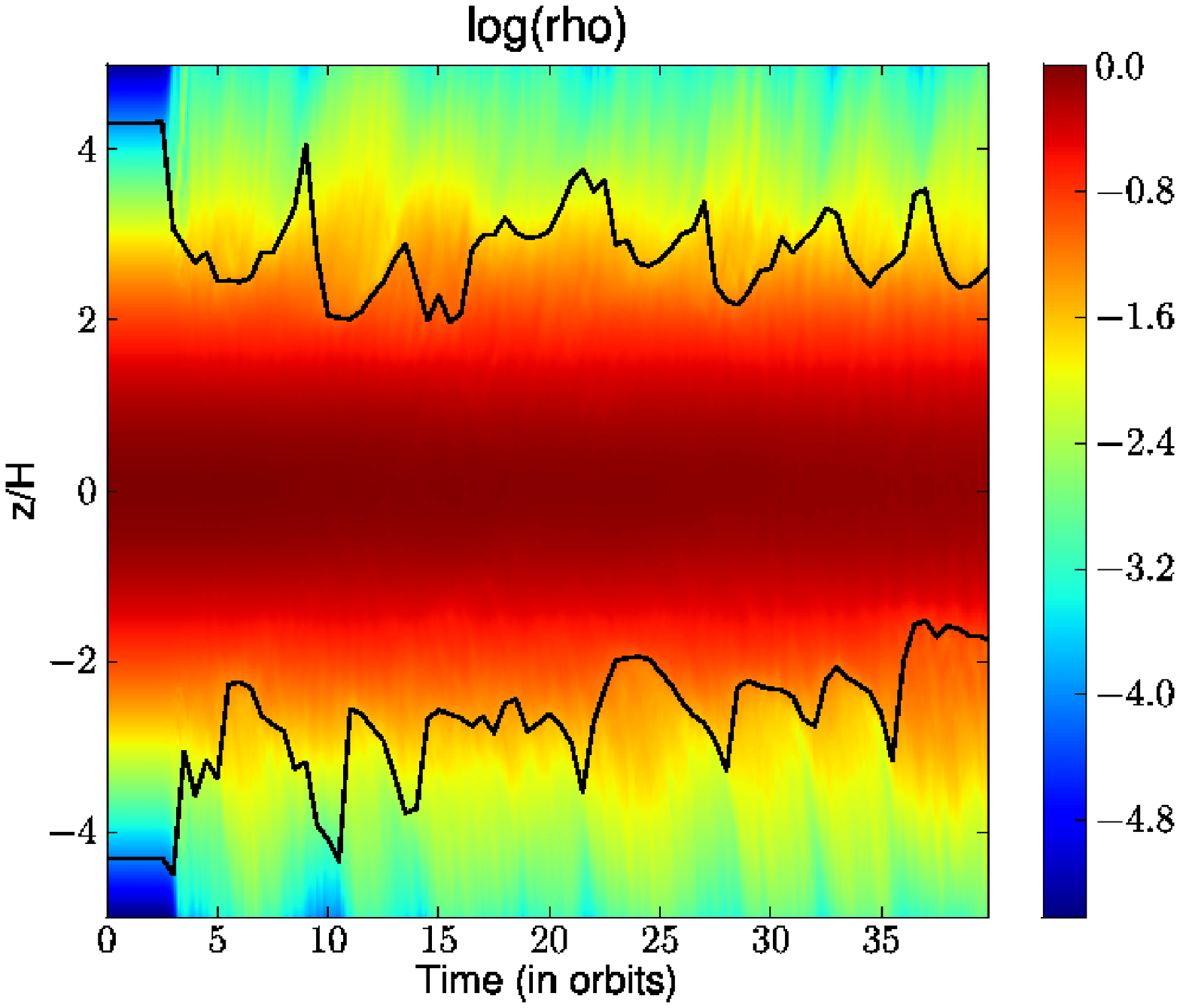}
\includegraphics[scale=0.43]{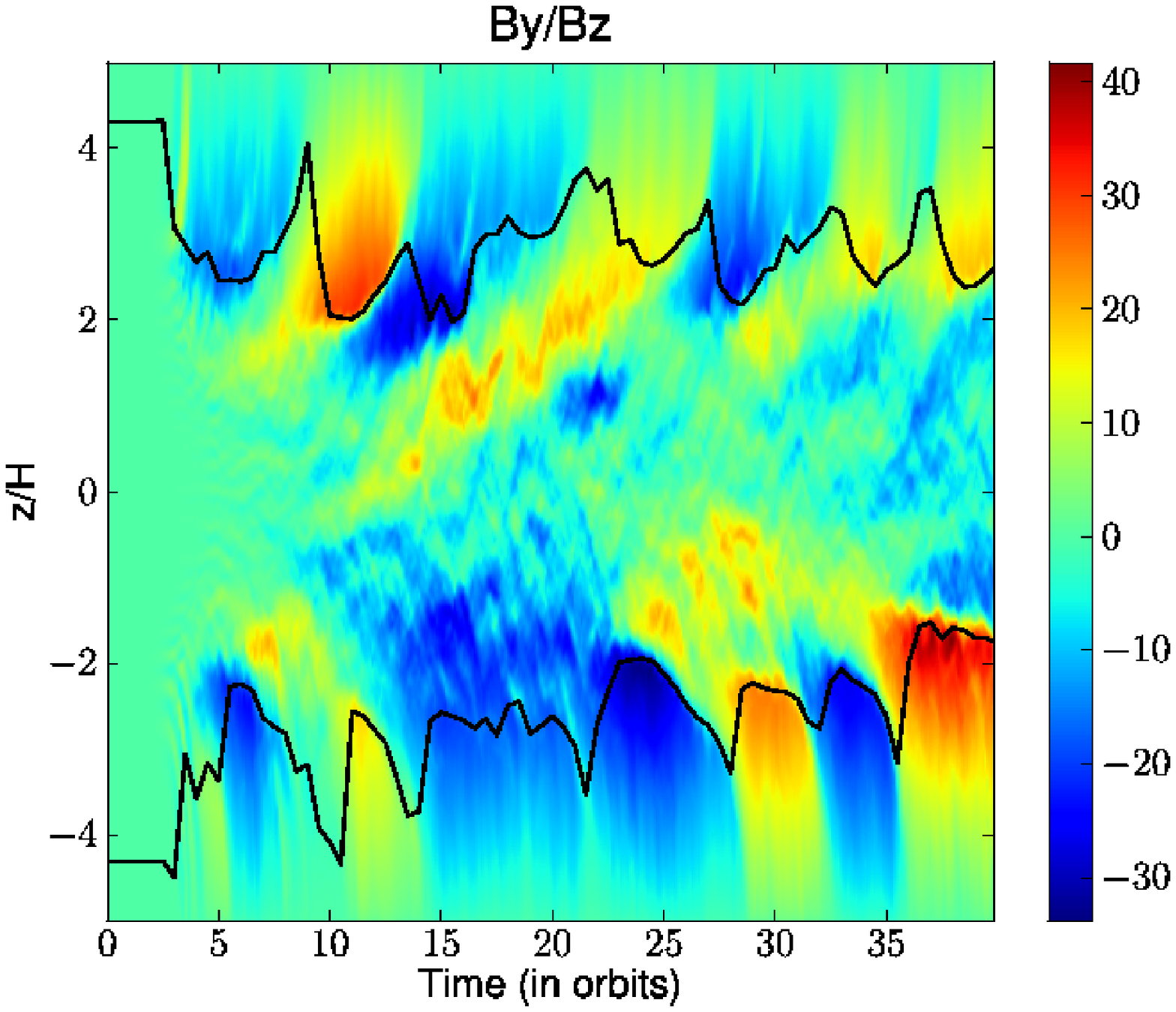}
\includegraphics[scale=0.43]{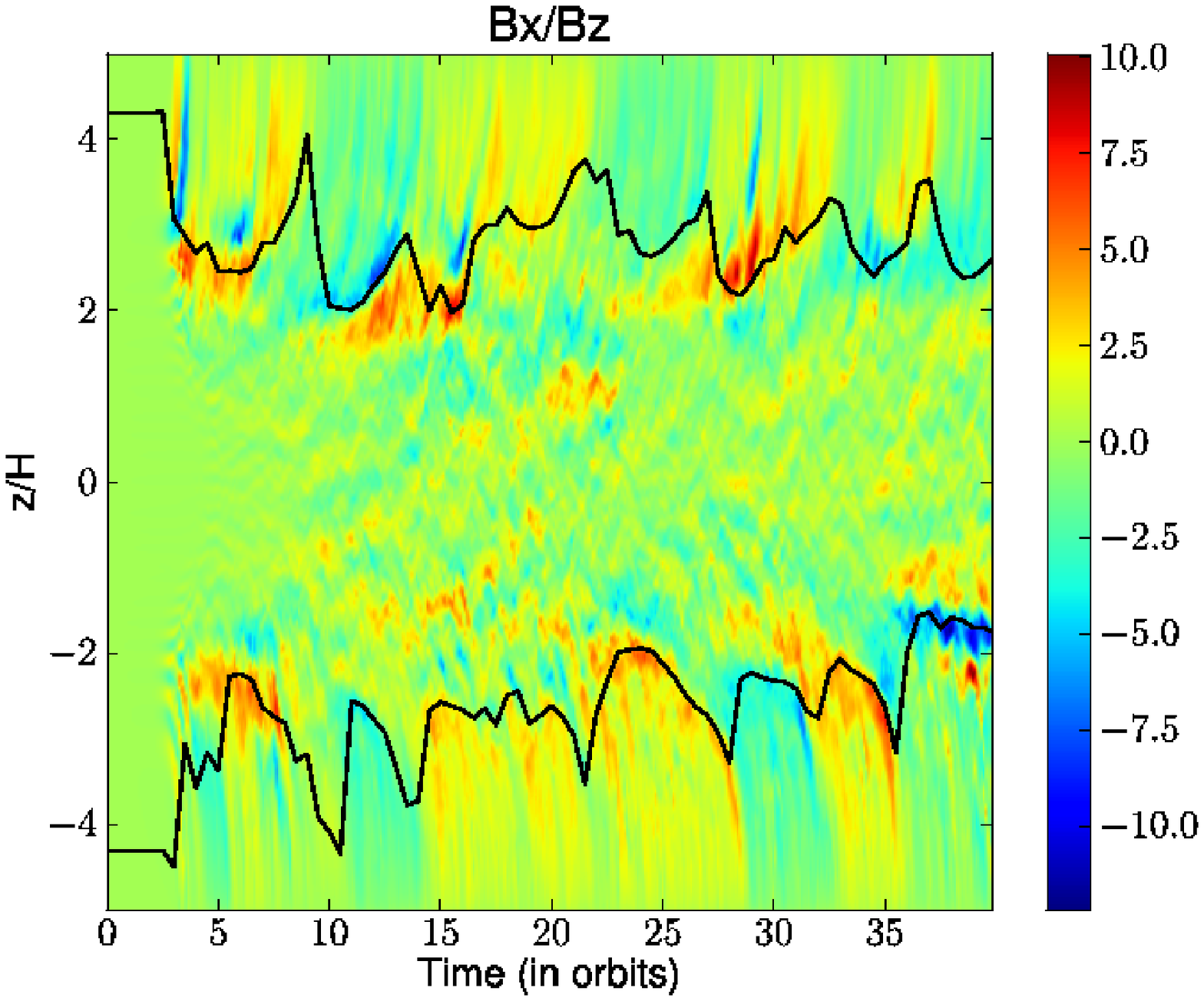}
\includegraphics[scale=0.43]{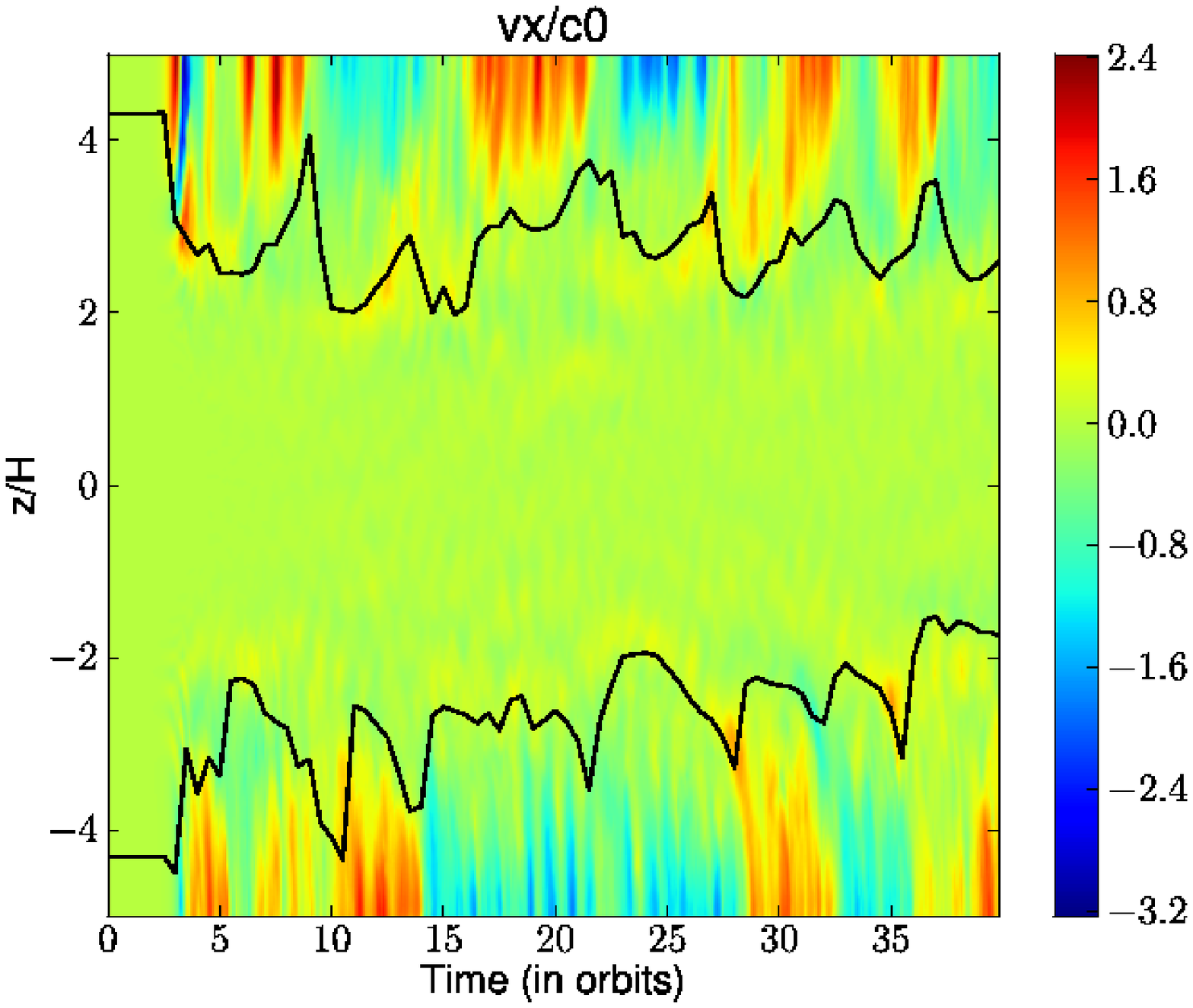}
\caption{Space-time diagram of the logarithm of the horizontally
averaged mass density ({\it upper left panel}), $\mean{B_y}/\mean{B_z}$ ({\it
upper right panel}), $\mean{B_x}/\mean{B_z}$ ({\it lower left panel})
and $\mean{v_x}/c_0$ ({\it lower right panel}) for model {\it
  Diffu4H}. For all panels, the solid line shows the location of unit
ratio of horizontally averaged thermal and magnetic pressures,
$\beta=1$.}
\label{spacetimeLx4h_fig}
\end{center}
\end{figure*}

We first illustrate the flow structure with the help of
figure~\ref{massFluxLx4h_fig} and \ref{spacetimeLx4h_fig}. The
first figure represents the time evolution of the mass loss rate
$\dot{m}_w$. On the second figure, four space-time
diagrams are plotted, representing the time-evolution of the vertical
profiles of various horizontally averaged quantities. In addition, the
location of unit ratio of the horizontally averaged thermal and
magnetic pressures, $\beta=1$, is overplotted on each panel with a
solid black line.

Several comments can be made based on these plots. First, the bursty
nature of the outflows is illustrated by
figure~\ref{massFluxLx4h_fig}: $\dot{m}_w$ displays quasi--periodic
variations during which the outflow rate increases by a 
factor of about $2$ to $3$. The typical duration of these bursts is
seen to be roughly $5$ orbits. In addition
figure~\ref{spacetimeLx4h_fig} shows that these bursts are associated
with higher density in the disk atmosphere and non-vanishing mean
magnetic fields and radial velocity. The fact that launching occurs
approximately at the location where $\beta=1$ is obvious from all the
panels. The toroidal component of the horizontally averaged magnetic
field $\mean{B_y}$ displays a cyclical behaviour above and below the
midplane, akin to the `butterfly diagram' reported by various authors  
\citep{gressel10,shietal10,davisetal10,simonetal11a} for stratified
zero-net-flux simulations. However, in contrast to the zero-net-flux
situation where the mean toroidal flux arises from the disk midplane,
it is seen here to originate at $z\sim 2$--$3H$, in agreement with the
results reported by SI09. As shown by figure\ref{massFluxLx4h_fig},
$\dot{m}_w$ is minimal when $\mean{B_y}$ changes sign, suggesting that
the mean magnetic field is involved in the launching mechanism. 

In the lower left and right panels, we plot the horizontally averaged
radial components of the magnetic field 
$\mean{B_x}$ (normalized by $\mean{B_z}$) and velocity $\mean{v_x}$
(normalized by the sound speed $c_0$) are respectively shown (Note
that, because of magnetic flux conservation, $\mean{B_z}$, which is
used to normalize $\mean{B_y}$ in the upper right panel, is
independent of $z$ and $t$ and equal to its initial value $B_0$). They
show that a significant mean radial magnetic field and 
mean radial velocity develop in the disk atmosphere. Their amplitudes are
comparable to those of $\mean{B_z}$ and $\mean{v_z}$, which means that
both the poloidal magnetic field lines and the poloidal streamlines
are significantly inclined with respect to the vertical axis. As
expected because of the shear, the radial and toroidal components of the
magnetic field are anti-correlated. Positive $\mean{B_y}$ corresponds to
negative $\mean{B_x}$ and vice versa. The radial velocity also correlates
positively with the radial magnetic field above the midplane
(i.e.\ positive $\mean{v_x}$ corresponds to positive $\mean{B_x}$) but correlates
negatively with it below the midplane (where positive $\mean{v_x}$ corresponds to
negative $\mean{B_x}$). This flow morphology closely resembles that of the
classical wind solution described by \citet[][hereafter
BP82]{blandford&payne82}. There are differences, though.  First,
in contrast to the steady-state BP82 solutions, the flow is strongly
time-dependent here. Second, because of the symmetries associated with
the shearing-box model, outflows are associated with poloidal magnetic
field lines inclined towards either positive or negative $x$. This is
different from the situation usually considered in cylindrical geometry in which
magnetic field lines are inclined away from the central object in
order to produce a successful outflow.  As noted by \citet{ogilvie12},
the shearing box is unaware whether the central object is located in
the direction of positive or negative $x$, and the local physics of
wind launching is symmetrical. In our shearing-box simulations, there
can even be situations where the poloidal field lines are 
directed towards positive $x$ above the disk midplane, but
towards negative $x$ below the midplane (see, for
example, the structure of the flow for $15<t<20$ in
figure~\ref{spacetimeLx4h_fig}). Such unfamiliar flow morphologies are
however permitted solutions of the equations in the framework of the
shearing-box model \citep{ogilvie12,lesuretal12}, and suggest that the
outflows above and below the midplane are to some extent independent
of each other. 

\begin{figure}
\begin{center}
\includegraphics[scale=0.43]{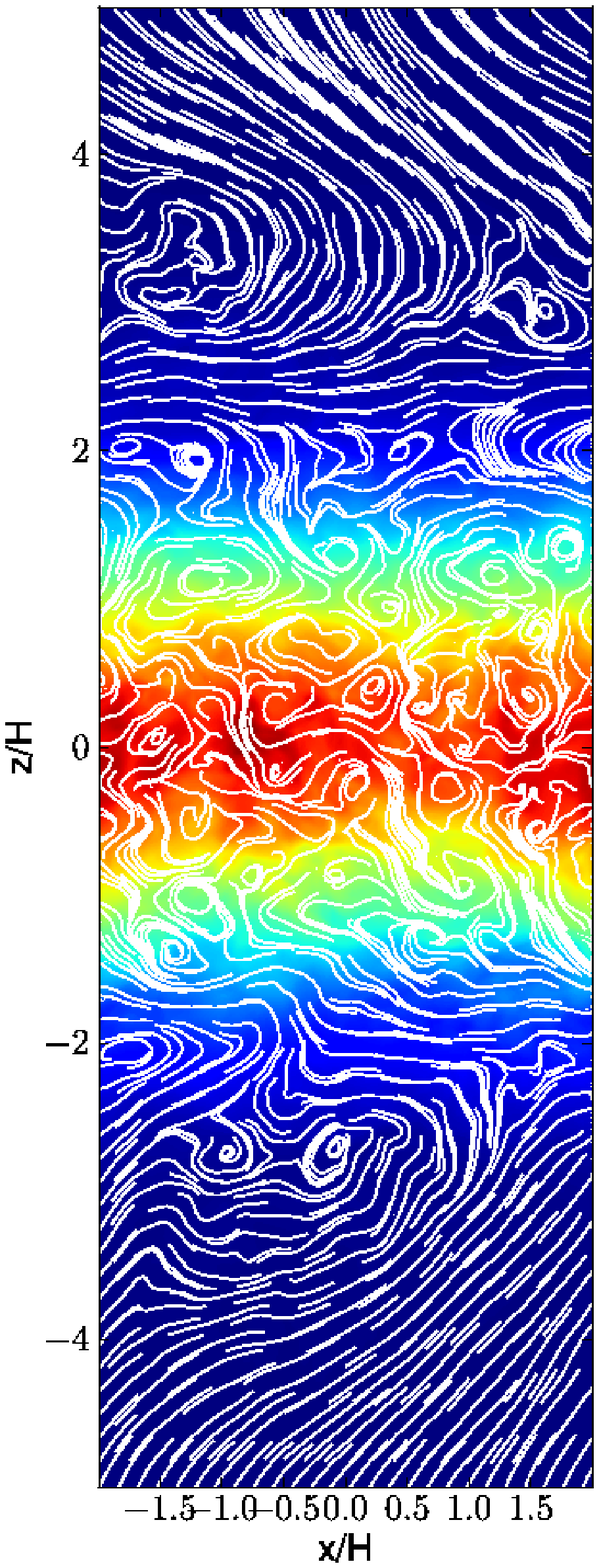}
\includegraphics[scale=0.43]{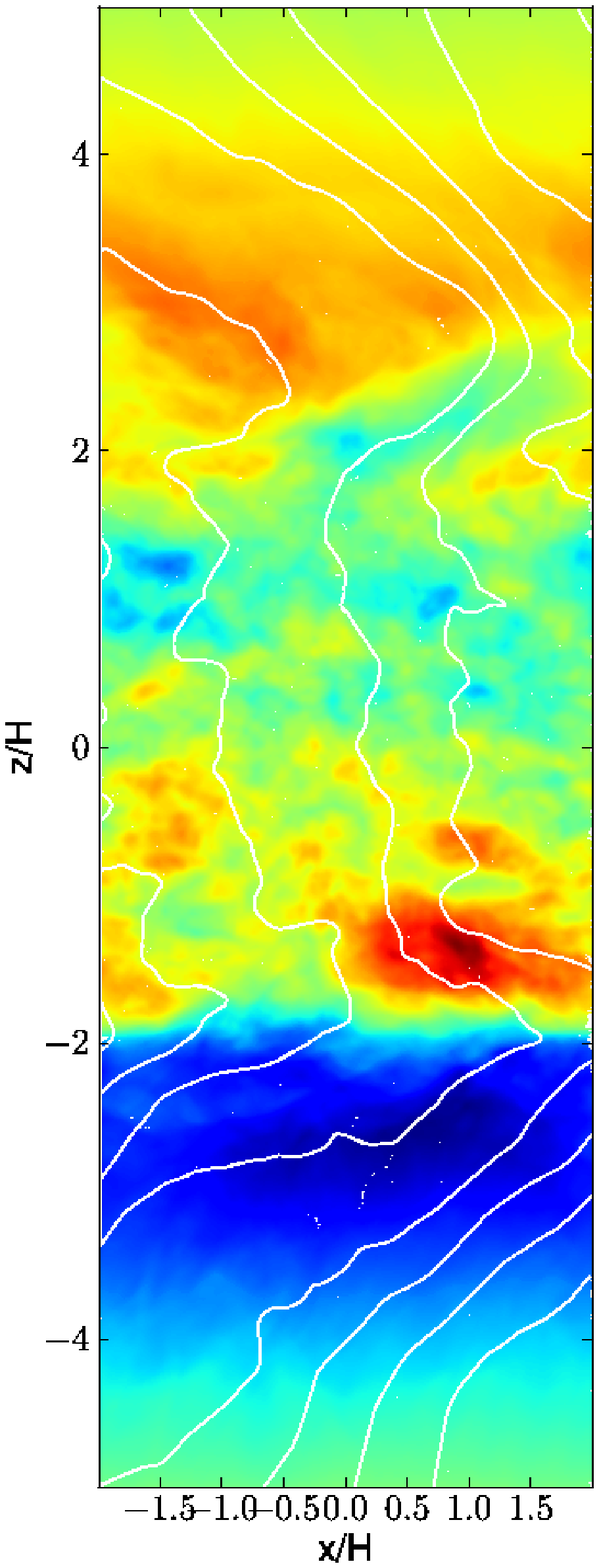}
\caption{$y$--averaged poloidal streamlines ({\it left panel}) and
  poloidal magnetic field lines ({\it right panel}) in the $(x,z)$
  plane, time--averaged over the ejection event occurring at
  $22<t<27$, and overplotted on the density distribution ({\it left
    panel}) and azimuthal magnetic field distribution ({\it right panel}).}
\label{lines_burst_fig}
\end{center}
\end{figure}

Because of this variety of field configurations, and since the flow is
not in steady state, it is not straightforward to illustrate the
analogy with the BP82 solutions more quantitatively. In attempting to
do so, we have isolated a single burst event that happens between the times
$t=22$ and $t=27$ and have analysed it in more detail. Time-averaging
the simulation data over such a short interval obviously means that
the resulting curves will be noisy, but this is the price to
be paid for a quantitative comparison with the theoretical
expectations. The structure of the time--averaged flow is illustrated
in figure~\ref{lines_burst_fig}. Poloidal streamlines and magnetic
field lines, spatially averaged in the $y$ direction and
time--averaged over the ejection event, are represented in the $(x,z)$
plane. They are overplotted on the density and azimuthal magnetic
field distribution respectively. Poloidal magnetic field lines are
systematically bent towards negative $x$ for $z \geq 3H$ and poloidal
streamlines tend to align with those magnetic field lines, in a way
that resembles Blandford \& Payne type of solutions. In addition, a
layer with a strong $B_y$ is apparent at the base the launching
region, again consistent with this picture. In the following, we
analysed quantitatively the properties of that flow.

\begin{figure}
\begin{center}
\includegraphics[scale=0.45]{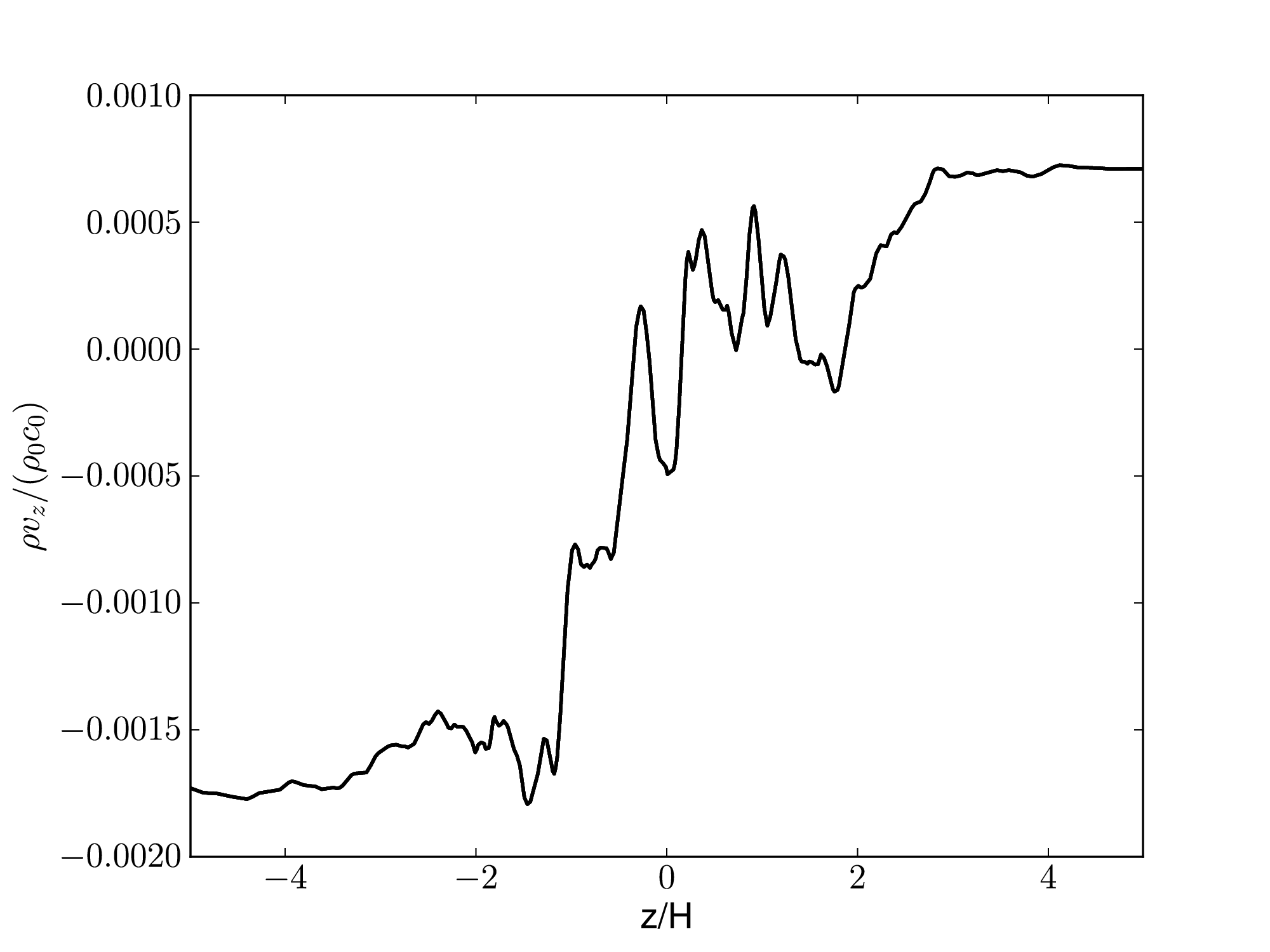}
\caption{Vertical profile of the horizontally averaged vertical mass flux,
  $\mean{\rho v_z}$, obtained in model {\it Diffu4H}. The data are time-averaged over a single burst between $t=22$ and $t=27$ and
  normalized by $\rho_0 c_0$.}
\label{massflux4h_fig}
\end{center}
\end{figure}

\begin{figure}
\begin{center}
\includegraphics[scale=0.45]{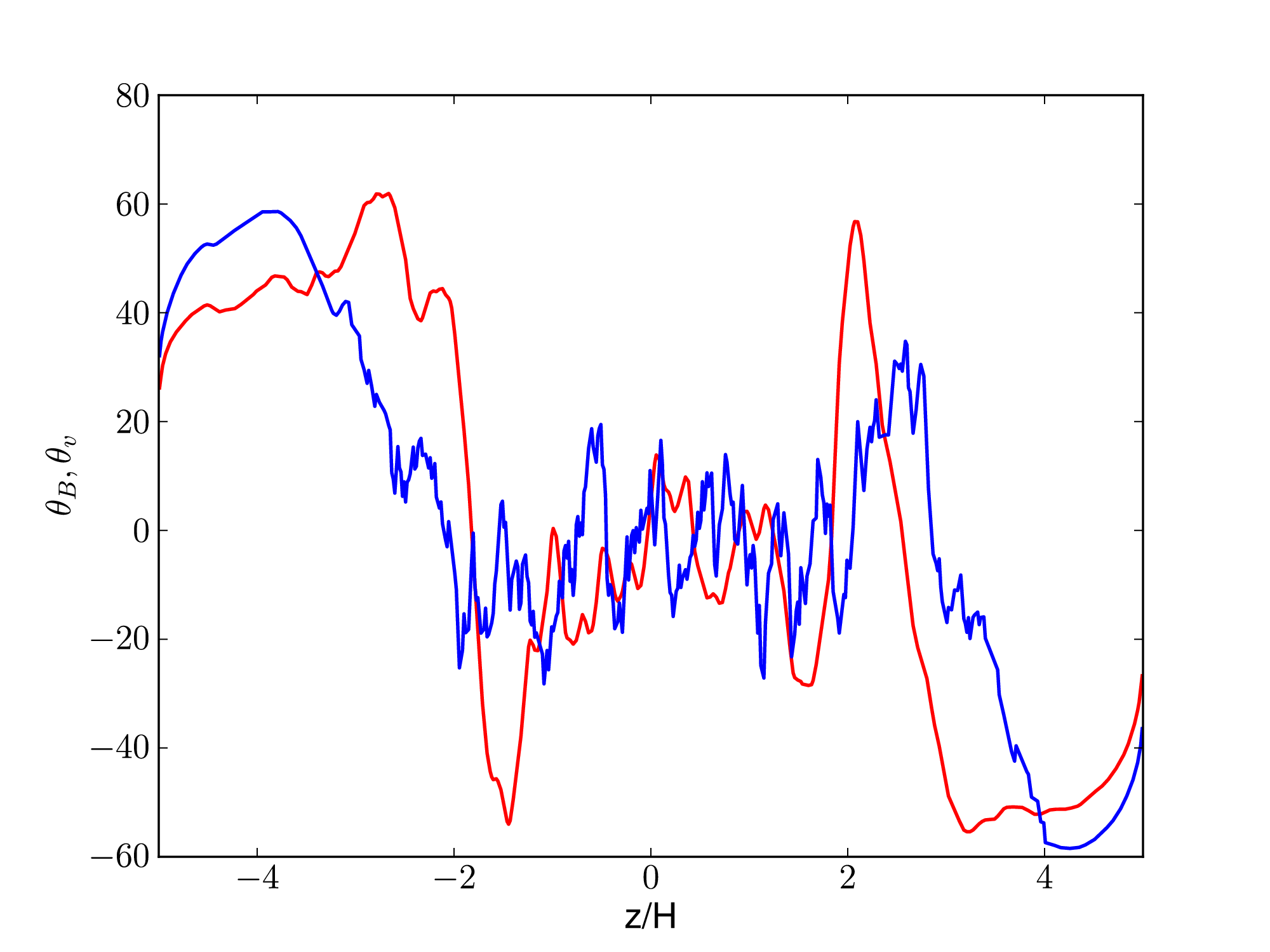}
\caption{Vertical profiles of the angles $\theta_B$ ({\it red curve})
  and $\theta_v$ ({\it blue curve}), measured between the vertical
  axis and the horizontally averaged poloidal magnetic and velocity
  fields, respectively. Both curves are obtained from model {\it
    Diffu4H} and are time-averaged over a single burst of ejection
  between $t=22$ and $t=27$.} 
\label{angles4h_fig}
\end{center}
\end{figure}

\begin{figure}
\begin{center}
\includegraphics[scale=0.45]{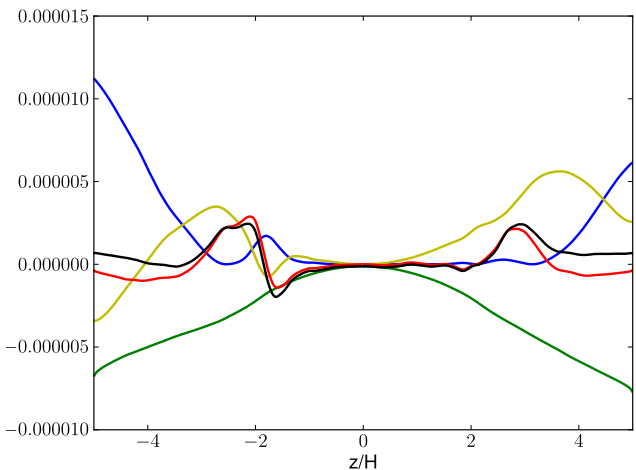}
\caption{Vertical profiles of the quantities $E_K$ ({\it blue
    curve}), $E_B$ ({\it red curve}), $E_T$ ({\it green curve}),
  $E_\phi$ ({\it yellow curve}) and their sum $E$ ({\it black}) along a
  particular magnetic field line. Data have been horizontally averaged
  and time averaged between $t=22$ and $t=27$, i.e. over a single
  ejection event.}
\label{bernouilli_fig}
\end{center}
\end{figure}

\begin{figure}
\begin{center}
\includegraphics[scale=0.45]{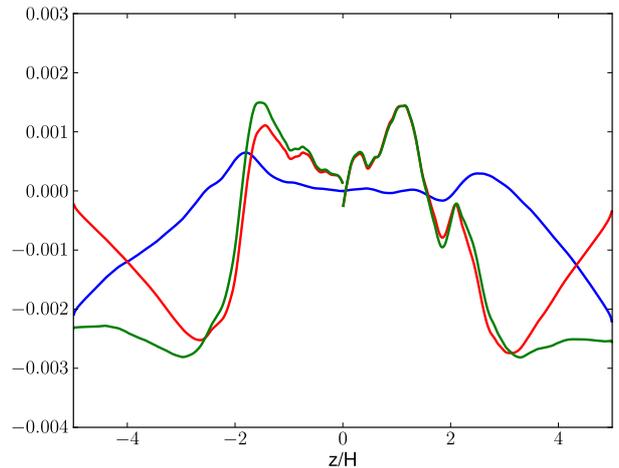}
\caption{Vertical profiles of the quantities $\mathcal{L}$ ({\it blue
    curve}), $\mean{B_y}/\alpha$ ({\it red curve}) and the expected invariant
  $f=\mathcal{L}-\mean{B_y}/\alpha$ ({\it green curve}) along a
  particular magnetic field line, computed as in
  figure~\ref{bernouilli_fig}.} 
\label{L0_fig}
\end{center}
\end{figure}

The first diagnostic we obtained is provided by the
vertical mass flux. Its vertical profile is shown in
figure~\ref{massflux4h_fig}. While small and fluctuating in the bulk
of the disk, $\mean{\rho v_z}$ reaches a constant value for $|z/H|
\geq 3$. This means the flow has reached a steady state in the disk
atmosphere, and we can write
\begin{equation}
\mean{\rho v_z}= \alpha \mean{B_z} \, ,
\end{equation}
where $\alpha$ is constant beyond $|z|\sim 3H$. Under such conditions,
it is well known that the flow is characterized by  
a number of features and invariants first studied in global models of
disk winds \citep[BP82;][]{pelletier&pudritz92} and recently extended to the
case of the shearing box \citep{lesuretal12}. First, poloidal
streamlines approximately align with poloidal magnetic field lines, as shown on
figure~\ref{lines_burst_fig}. In figure~\ref{angles4h_fig}, we plot
the vertical profiles of $\theta_B$ and $\theta_v$, which respectively
denote the angles between the vertical axis and the horizontally averaged
poloidal magnetic and velocity fields. In those regions where the vertical mass
flux is constant, both $\theta_B$ and $\theta_v$ reach significant and
fairly $z$-independent values of order $60^\circ$. These angles are
comparable with each other ($\theta_B \sim \theta_v$) and
larger than $30^\circ$, which is the critical angle in the BP82
analysis. This supports the BP82 scenario: poloidal streamlines are
approximately aligned with poloidal magnetic fields and
magneto-centrifugal launching is possible. Investigating whether
this is actually the case can be examined using the Bernoulli
invariant, which should be constant along magnetic field lines
\citep{lesuretal12}: 
\begin{eqnarray}
E_{Ber} &=&E_K+E_T+E_\phi+E_B \nonumber \\
&=& \frac{\mean{v}^2}{2}+c_0^2
\log(\mean{\rho})+\phi-\frac{\mean{B_y}v_y^\star}{\alpha} \, ,
\label{bernouilli_eq}
\end{eqnarray}
where $v_y^\star=\mean{v_y}-\alpha\mean{B_y}/\rho$ and $\phi$
represents the shearing box tidal potential. The variations of the four
components of $E_{Ber}$ along a particular field
line originating at the disk midplane are plotted in
figure~\ref{bernouilli_fig}. Their sum is also represented using a
black line, and exhibits a relatively flat profile beyond $|z|\sim 3H$,
consistent with the 
expectations of steady state theory. Acceleration of the outflow, as
seen by an increase of kinetic energy, occurs after the flow
has reached a maximum in the potential energy. Its acceleration
is essentially a result of a decrease of the potential energy along the
field line. This demonstrates that the outflows are centrifugally
driven, as observed for global simulations of laminar resistive disks. Note also that the variations of the thermal energy $E_T$
contribute to about one third of that acceleration, indicating a
non--negligible effect of thermal pressure even above the Alfven
point. This importance of thermal pressure is due to the plasma
$\beta$ being much larger in the disk atmosphere (it is of order $1$)
than is common in standard studies of outflow launching (where usually
$\beta \ll 1$ in the launching region). 

It is well known that such disk outflows carry angular momentum away
from the disk midplane: angular momentum is extracted from the 
disk by the magnetic field before being transferred to gas as it is
accelerated \citep{lesuretal12}. This exchange is described by the following
invariant that should be constant on streamlines:
\begin{equation}
f=\mathcal{L}-\frac{\mean{B_y}}{\alpha},
\end{equation}
where $\mathcal{L}=\mean{v_y}+(2-q)\Omega x$ plays the role of specific angular momentum in the shearing-box approximation. Its vertical profile
along a particular streamline is plotted in figure~\ref{L0_fig}. Its
components $\mathcal{L}$ and $\mean{B_y}/\alpha$ are also represented
for comparison. Beyond $|z|=3H$, $f$ shows a flat
profile, with weak variations that are much smaller than the
variations of its constituents. This is again consistent with the
expectations of BP82: disk outflows extract angular momentum from the
disk. As shown by \citet{lesuretal12}, this leads to radial advection
of gas and poloidal magnetic field lines. However, we failed to
detect such an advection velocity. Averaging the radial mass flux
over the single burst, we found 
\begin{equation}
v_x^*=\frac{\int\int \mean{\rho v_x} dtdz}{\int\int \mean{\rho}dtdz}=2 \times 10^{-3} c_0
\end{equation}
This value is much below the turbulent velocity fluctuations, for
which the Mach number is at the $10\%$ level. This suggest 
that outflow-mediated angular momentum transport is very small, as
expected given the large value of $\beta_0$ we use, and consistent
with the low altitude of the Alfv\'en point. We shall return into this 
issue in the discussion. We note also that even if the radial
mass flux associated with these bursts appears to be small, it 
might still be locally important and influence the radial transport
of vertical magnetic flux, as suggested by
\citet{spruit&uzdensky05}.

While enlightening as to the structure of the flow in the disk
atmosphere, the above analysis should not hide the fact that the
outflows that develop in the simulations are not strictly speaking 
BP82-type disk winds. First, one should not forget the
non-steady nature of the flow, the properties of which are yet to be
understood. How much of that variability is an artifact of the
shearing-box symmetries should be investigated further. Another
important difference 
with classical disk wind theories is that the outflows presented here
do not pass all the relevant critical points within the computational domain.
It is unlikely that the successful complete ejection of the outflow
can be demonstrated within the local shearing-box model because the
depth of the gravitational potential is not represented. We shall
return to that point in the discussion (section~\ref{smooth_pot_sec}). To
summarize, even if the comparison drawn above is appealing, outflows
driven by turbulent disks are clearly more complex than described by
simple steady-state disk wind models. Working out simple models that
will simultaneously describe the midplane turbulent flow and the
non-steady BP82-like winds are beyond the scope of this 
paper and will surely prove challenging in the future.

\subsection{Disk midplane: turbulent  transport}
\label{turb_transport_sec}

\begin{figure}
\begin{center}
\includegraphics[scale=0.45]{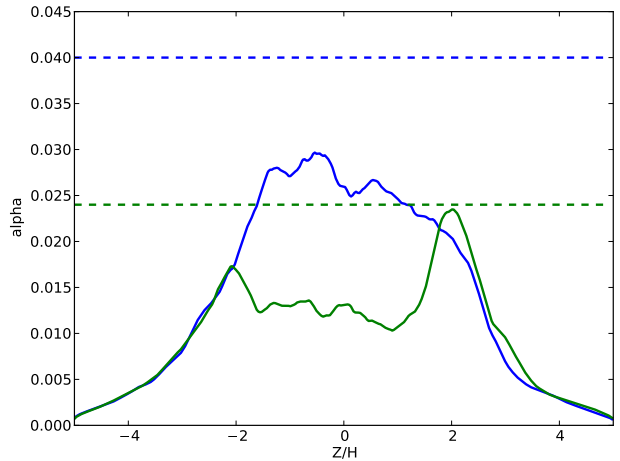}
\caption{Vertical profile of the total stress tensor, normalized by
the midplane pressure, for models {\it Diffu4H} ({\it blue line}) and 
{\it Diffu4HPmhalf} ({\it green line}). The blue and green dashed lines
represent the corresponding values obtained in unstratified runs
having the same parameters.}
\label{alphaVsZ_fig}
\end{center}
\end{figure}

In this section, we focus on angular momentum transport. As argued
above (see also section~\ref{wind_transport_sec}), it is likely
dominated by turbulent transport occurring in the bulk of the
disk. Here, we thus 
present a preliminary study of the saturation properties of the
turbulence in stratified shearing boxes threaded by a 
mean magnetic field. In addition, we compare our results with
identical simulations performed in the unstratified limit in order to
assess the effect of density stratification on transport
properties. Given the large computational expense associated with
such a survey, we defer a systematic study of the scaling with
vertical field amplitude to a future publication and consider here only the
effect of the dissipation coefficients. To do so, we consider two models,
the parameters of which are respectively $Re=Rm=3000$ (model {\it
Diffu4H}) and $(Re,Rm)=(3000,1500)$ (model {\it Diffu4Ha}). Thus
$Pm=1$ in the former and $0.5$ in the latter.

The $xy$-component of the sum of the Reynolds and Maxwell stresses,
normalized by the midplane pressure, can be taken to define a
local value for the Shakura--Sunyaev $\alpha_{SS}$ viscosity parameter. 
Its vertical profile is plotted in figure~\ref{alphaVsZ_fig} for both models. As
reported in the literature for different configurations, turbulent transport
increases with $Pm$. We found the vertically averaged value to be
$\alpha_{SS}=6.2 \times 10^{-2}$ when $Pm=1$ and $\alpha_{SS}=4.8 
\times 10^{-2}$ when $Pm=0.5$. The vertical profile of the stress we
obtained in each case is comparable to that reported in the literature
by many authors using local or global simulations
\citep{miller&stone00,hiroseetal06,fromang&nelson06,flaigetal10,nataliaetal10,sorathiaetal10,fromangetal11,
simonetal11a}. It is rather 
flat in the bulk of the disk, with possible local maxima at $z \sim
2H$ before dropping to low values in the corona. 
Theoretical arguments tentatively explaining such a profile have
also recently been proposed in the literature
\citep{guan&gammie11,uzdensky12}. The reason why the local maxima are
more apparent in the lower $Pm$ case is unclear and deserve future
investigations, but might be related to the different turbulent states
described by \citet{simonetal11a} for vanishing vertical flux
configurations or simply to the resistivity being larger in the case
$Pm=0.5$. As shown by the dashed 
lines in figure~\ref{alphaVsZ_fig}, the transport rates we
measured are in acceptable agreement with results obtained in
unstratified boxes. In such cases, we indeed obtained $\alpha_{SS}=4.0 \times
10^{-2}$ and $2.4 \times 10^{-2}$ when $Pm=1$ and $0.5$,
respectively. These values are higher by about a factor of two than the 
midplane value of the transport rates obtained in the stratified
  models and display the same trend with magnetic Prandtl
number. This means that unstratified shearing-box simulations with net
flux, such as reported by \citet{longaretti&lesur10}, provide a good
first-order estimate of angular momentum transport in stratified
disks. 

\section{Discussion}
\label{discussion_sec}

In this section, we discuss several controversial aspects of the
present simulations, namely the influence of vertical boundary
conditions on our results, the relative importance of turbulent
and outflow-mediated angular momentum transport, and possible issues
associated with a straightforward truncation of the shearing-box
vertical potential.

\subsection{Influence of the vertical boundary conditions}

\begin{figure}
\begin{center}
\includegraphics[scale=0.45]{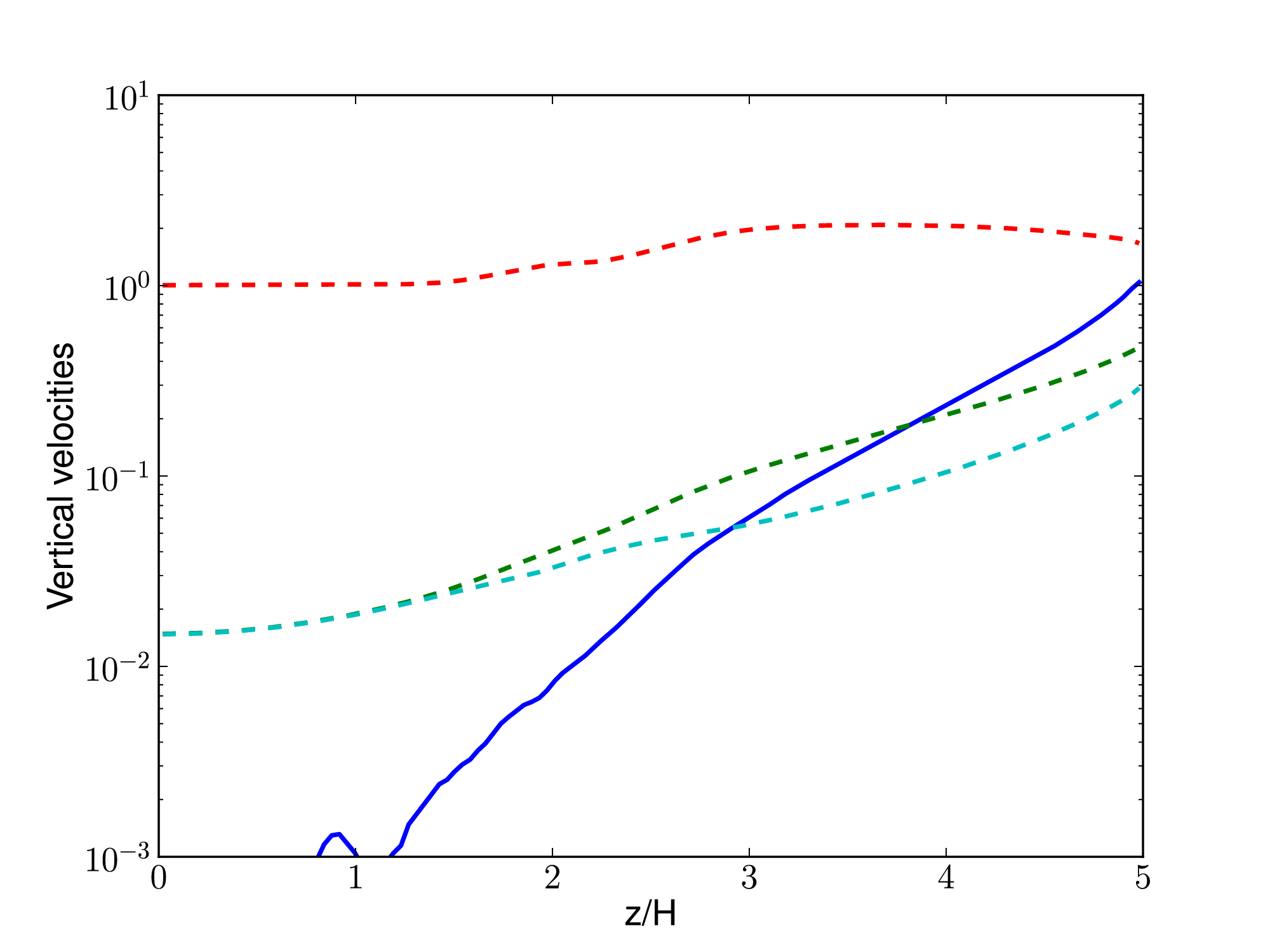}
\caption{Same as figure~\ref{sonicPoints_fig}, but for the model
having zero-gradient vertical boundary conditions on $B_x$ and $B_y$.}
\label{sonicPoints_zerograd_fig}
\end{center}
\end{figure}

Clearly, it is not satisfying to see such a strong dependence of the
outflow mass-loss rate on numerical aspects of the setup. One might
worry that some of this dependence is due to the vertical boundary
conditions. In particular, our choice of purely vertical magnetic
fields at the top and bottom of the computational box is rather
unfortunate. Indeed, it does not favour magneto-centrifugal
acceleration as the latter requires magnetic field lines inclined by
more than $30^{\circ}$ to operate. This might affect the flow and
prevent successful launching of disk outflows.

In order to test that possibility, we performed an additional
simulation with different boundary conditions on the magnetic
field. We adopted zero-gradient boundary conditions for $B_x$ and
$B_y$ at the top and bottom of the box. To save computational power,
this simulation was performed without explicit dissipation
coefficients and with a smaller resolution of $32$ cells per $H$, using
the same box size as for model {\it Diffu4H}, namely
$(L_x,L_y,L_z)=(4H,4H,10H)$. We obtained $\dot{m}_w=1.0 \times
10^{-3}$. This is very similar to the mass outflow rates obtained for
models {\it Diffu4H} and {\it Diffu4Ha} that share the same box
size. As an additional check, we plot in
figure~\ref{sonicPoints_zerograd_fig} a comparison between the
vertical gas velocity and the various wave velocities of the problem,
as done for model {\it Diffu4H} in the top panel of
figure~\ref{sonicPoints_fig}. The two plots are very similar,
indicating a weak effect of the boundary conditions on the flow
properties (note, however that, as opposed to mode {\it Diffu4H}, the
flow velocity remains smaller that the sound speed everywhere).

Thus, we have shown in this paper that our results are unchanged when
we adopt widely used boundary conditions for the magnetic field
(purely vertical field or zero gradient for its horizontal
components). The good agreement with the results published by SI09 and 
SMI10 suggests that this insensitivity to vertical boundary conditions
extends to more elaborate boundary conditions. This is strange as
we have shown that the results depend on vertical box size
(section~\ref{Lz_sec}) which indicates that the flow knows about the
existence of an upper boundary. Its seems that the artifacts uncovered
in the present paper are due to the very existence of a boundary rather
that the nature of the conditions imposed there. Alternatively, this
could indicate a breakdown of the shearing--box model itself. Future
investigations along with comparison with global simulations are
needed to clarify this issue.


\subsection{Vertical vs. turbulent transport of angular momentum}
\label{wind_transport_sec}

\noindent
The paradigm that emerges out of the analysis performed in this paper
is that of a turbulent disk surrounded by transient and presumably
local magneto-centrifugally accelerated BP82-like outflows. As demonstrated in
figure~\ref{L0_fig}, such outflows carry angular momentum away from the
bulk of the disk, the amount of which might be important for its
evolution. In this section, we speculate on the relative importance of the
angular momentum transported vertically by the outflow and radially by the
turbulence.

Given the symmetries of the shearing box, there is some ambiguity in
determining that ratio. One possibility we explored is to separate the
flow into mean fields and fluctuations and to assign the transport
mediated by the former to the outflow and that due to the latter to the
turbulence. However, 
we found significant radial transport due to the mean magnetic field,
the role of which remains unclear. We therefore followed the approach
sketched by \citet{wardle07} for a radially structured accretion disk. In
cylindrical coordinates $(R,\phi,Z)$, we first write the horizontal
velocities $v_R$ and $v_{\phi}$ as the sum of an azimuthally averaged
part and remaining fluctuations: 
\begin{eqnarray}
v_R &=& \mean{v_R}+\delta v_R \, , \\
v_{\phi} &=& \mean{v_{\phi}}+\delta v_{\phi}=R\Omega + \delta v_{\phi} \, .
\end{eqnarray}
Angular momentum conservation in a steady-state disk is then written
\begin{equation}
\label{ang_mom_turb_eq}
R \mean{\rho v_R}\frac{\partial R^2 \Omega}{\partial R} =
\frac{\partial}{\partial R} \left( R^2 T_{R \phi} \right) +
\frac{\partial}{\partial Z} \left( R^2 T_{Z \phi} \right) \, ,
\end{equation}
with the turbulent stress tensors $T_{R \phi}$ and $T_{Z \phi}$
expressed as \citep{balbus&pap99}:
\begin{eqnarray}
T_{R \phi} &=& \mean{B_R B_{\phi}-\rho \delta v_R \delta v_{\phi}} \label{rphi_turb} \\
T_{Z \phi} &=& \mean{B_Z B_{\phi}-\rho v_Z \delta v_{\phi}} \, , \label{zphi_turb}
\end{eqnarray}
We now assume that there exist disk surfaces located at $Z=\pm Z_s(R)$
above and below which mass and angular momentum are
outflowing. We define the mass accretion rate in the disk as
\begin{equation}
\dot{M}=-2 \pi R \int_{-Z_s}^{+Z_s} \mean{\rho v_R} dZ
\end{equation}
and integrate Eq.~(\ref{ang_mom_turb_eq}) between $-Z_s$
and $+Z_s$. We find 
\begin{equation}
\frac{\dot{M}}{2 \pi R} \frac{\partial R^2 \Omega}{\partial R} = \dot{m}_R +
\dot{m}_Z \, ,
\end{equation}
where $\dot{m}_R$ and $\dot{m}_Z$ respectively denote the disk and the outflow
contribution to the angular momentum transport and are 
given by
\begin{eqnarray}
\dot{m}_R &=& - \frac{1}{R} \frac{\partial}{\partial R} \left( R^2 \int_{-Z_s}^{+Z_s} T_{R
    \phi} dZ \right) \, , \label{fr_eq} \\
\dot{m}_Z &=& - R \left[ T_{Z\phi} \right]_{-Z_s}^{+Z_s}  \, . \label{fz_eq}
\end{eqnarray}
The relative importance of turbulence and outflowing material to
angular momentum transport is given by the ratio of $\dot{m}_R$ and
$\dot{m}_Z$. In the shearing box, however, $\dot{m}_R$ artificially
vanishes because of symmetries of the model. It is possible, though,
to use the shearing box results to make such a evaluation for real
disks, provided reasonable assumptions are made about the disk
structure. Let us first assume 
that the vertical profile of the stress, when normalized by the
midplane pressure $P_{mid}$, is a   ``universal'' function of $\tilde
Z=Z/H$ only, and is independent of $R$ (which amounts to saying that
$\alpha$ is radially constant). Then all 
the radial dependence in $\dot{m}_R$ can be made explicit 
provided we give ourselves radial scalings for the surface density
$\Sigma \propto R^p$ and the sound speed $c_0^2 \propto R^q$ (we
assume a locally isothermal equation of state, i.e. that $c_0$ a function
of $R$ only, which means the vertical density
profile is Gaussian). After some algebra, we obtain
\begin{equation}
\dot{m}_R= (p+q+2) \overline{\alpha_R} \frac{\Sigma c_0^2}{\sqrt{2 \pi}},
\end{equation}
where we have defined the radius-independent variable $\overline{\alpha_R}$ as
\begin{equation}
\overline{\alpha_R}=\int_{-\tilde{Z_s}}^{+\tilde{Z_s}} \alpha_R(\tilde Z) d\tilde{Z}=-\int_{-\tilde{Z_s}}^{+\tilde{Z_s}} \frac{T_{R\phi}}{P_{mid}} d\tilde{Z},
\end{equation}
and the second equality serves as a definition
of $\alpha_R(\tilde Z)$. The interest of the last pair of equations is that
$\overline{\alpha_R}$ can be derived from shearing-box simulations
such as presented in the present paper. Similarly, $\dot{m}_Z$ can be written as
\begin{equation}
\dot{m}_Z= \left( \frac{R}{H} \right) [\alpha_Z]_{-Z_s}^{+Z_s} \frac{\Sigma c_0^2}{\sqrt{2 \pi}},
\end{equation}
where $\alpha_Z=-T_{Z\phi}/P_{mid}$, so that the ratio
$\dot{m}_R/\dot{m}_Z$ is given by 
\begin{equation}
\frac{\dot{m}_R}{|\dot{m}_Z|}=(p+q+2) \left( \frac{H}{R} \right)
\frac{\overline{\alpha_R}}{\left|[\alpha_Z]_{-Z_s}^{+Z_s}\right|} \, .
\label{ratio_eq}
\end{equation}
Note the absolute value in the expression above which accounts for the
facts that outflow-mediated angular momentum transport can be directed
either way in the shearing box. We are interested here only in its
amplitude.

\begin{figure}
\begin{center}
\includegraphics[scale=0.45]{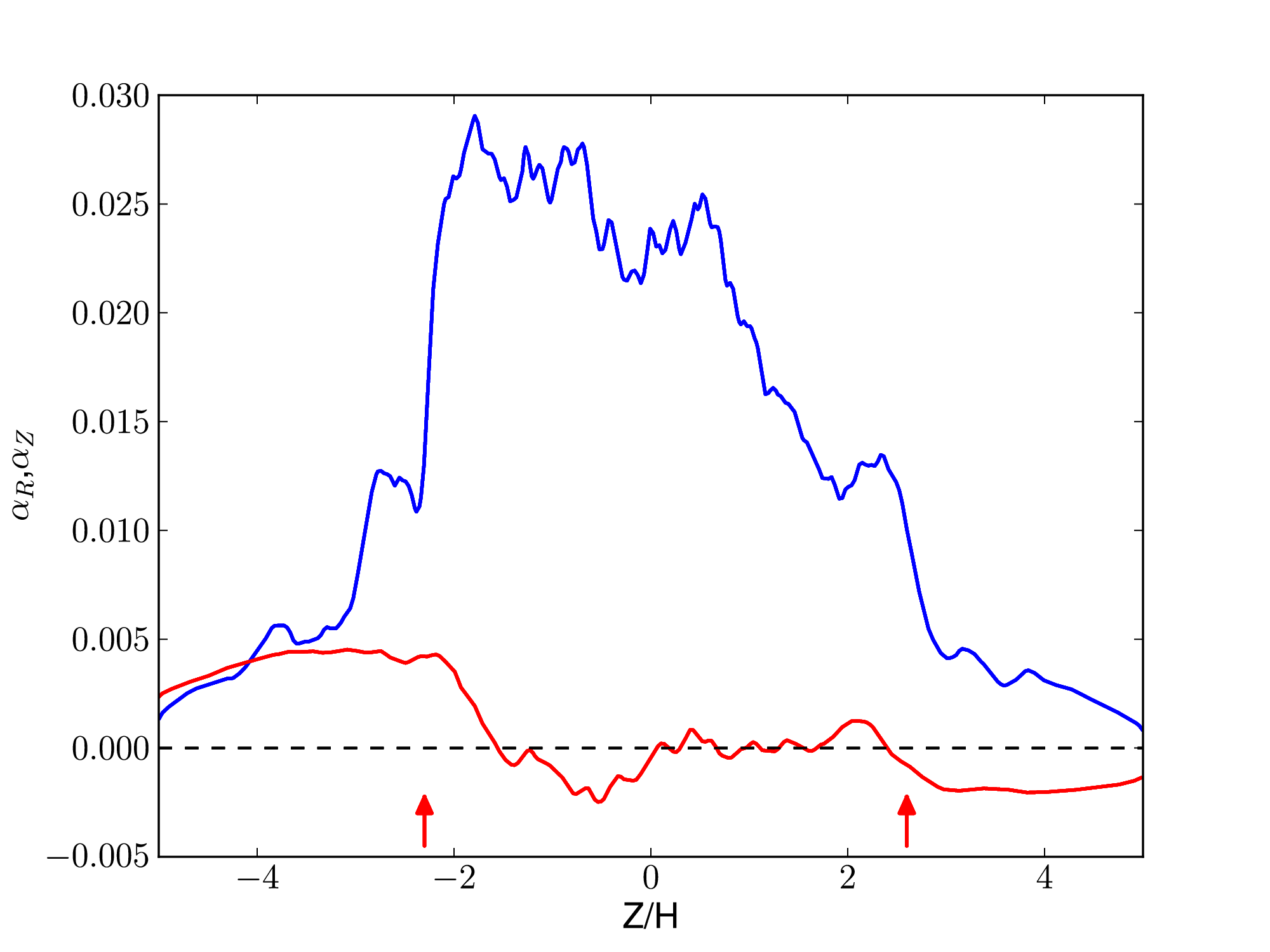}
\caption{Vertical profile of $\alpha_R$ ({\it blue curve}) and
  $\alpha_Z$ ({\it red curve}), time averaged using the data of model
  {\it Diffu4H} over the interval $22<t<27$. Vertical arrows marks the
  location of unit ratio between horizontally averaged thermal and
  magnetic pressures.} 
\label{angmom_flux_fig}
\end{center}
\end{figure}

Using model~{\it Diffu4H}, we measured $\alpha_R$ and $\alpha_Z$. We
focus on the same burst that we studied in
section~\ref{outflow_struc_sec}, and thus averaged the simulation data 
between $t=22$ and $t=27$. The vertical profiles we obtained by doing
so are plotted in
figure~\ref{angmom_flux_fig} along with vertical arrows showing the
ejection region calculated as in fig.~\ref{betaVsZ_fig}. At first
glance, fig.~\ref{angmom_flux_fig} shows than $\alpha_R$ is larger
than $\alpha_Z$. This is confirmed by precise measurements, which give
$\overline{\alpha_R}=1.0 \times 10^{-1}$ and
$[\alpha_Z]_{-Z_s}^{+Z_s}=-4.4 \times 10^{-3}$. Note that we have
taken $Z_s=2.5H$ in making that estimate. Taking $p+q+2$ to be of
order unity in Eq.~(\ref{ratio_eq}), this gives
\begin{equation}
\frac{\dot{m}_R}{|\dot{m}_Z|} \sim 24 \left( \frac{H}{R} \right) \, .
\end{equation}
This means that turbulent transport dominates outflow-mediated
transport as long as $H/R>0.04$. We stress, however, that outflows are
more difficult to launch in thinner disks because the escape
Mach number is larger in that case. Indeed, $v_{esc}=\sqrt{2}
v_{\phi}$, which translates into $v_{esc}/c_0=\sqrt{2} (R/H)$. Thus,
in thin disks where we can imagine outflow-mediated transport to
dominate, its vertical Mach number should exceed $\sim 35$ for
successful ejection. As demonstrated in both panels of
figure~\ref{sonicPoints_fig}, this is far from being the case in the
simulations we have been able to run, in which the vertical Mach
number reaches at most values of a few. Thus the question of whether
such outflows will eventually escape the potential well or fall back
on the disk is still an open issue. 

Finally, we end this section by stressing that the nature of the
transport in the disk corona is still uncertain. Although part of it 
should be attributed to the presence of the outflow, some of the
radial transport could be due to coronal magnetic loops transmitting
angular momentum between their footpoints as suggested by 
\citet{uzdensky&goodman08}. Since the disk outflow also carries
angular momentum radially in the shearing box model, it is difficult
to disentangle the relative contribution of the two mechanisms.

\subsection{Smoothing the shearing-box potential?}
\label{smooth_pot_sec}

It is obvious that the shearing-box model suffers from a number of
artifacts that might affect the results presented here. Maybe
the most important of them is the singular nature of the tidal
potential used in our simulations. In standard shearing-box simulations,
the potential is given as
\begin{equation}
\phi_{sb}=-\frac{3}{2}\Omega^2 x^2 + \frac{1}{2} \Omega^2 z^2 \, .
\label{phisb_eq}
\end{equation}
The second term shows that the potential becomes deeper and deeper as
the vertical box size is increased. This means it requires larger and
larger energy for gas to climb out of that potential, while in reality
the gravitational potential has a finite depth
\begin{equation}
\phi_g=-\frac{GM_{\star}}{\sqrt{R^2+Z^2}} \, .
\label{phigrav_eq}
\end{equation}
The second term in Eq.~(\ref{phisb_eq}) is simply the second-order
Taylor expansion of the last equation with respect to $z$.

SMI10 investigated this issue by directly employing the potential
given by Eq.~(\ref{phigrav_eq}). They obtain results that are
difficult to interpret as $\dot{m}_w$ display variations by factors of
$2$--$3$ and show no systematic trend with vertical box size. Here, we
caution that a more rigorous and self--consistent approach is to
include the next order in the expansion of the tidal potential. Up to
the first terms that include a modification of its vertical profile,
such an expansion is written
\begin{equation}
\phi = \phi_{sb} + \left( \frac{H}{R} \right) c_0^2
\left[ \tilde{x}^3 - \frac{3}{2} \tilde{x}\tilde{z}^2 \right]
\end{equation}
where we have noted $\tilde{x}=x/H$ and $\tilde{z}=z/H$ and use the
relation $H=c_0/\Omega$.

The expression above shows that the first additional terms in the
expansion involves $x$ as well as $z$. This means that smoothing the
potential should be done by simultaneously altering the radial
component of the tidal force. We note that in such a situation,
curvature terms should also be accounted for, and the
shearing-periodic boundary conditions would be incompatible with the
governing equations. The first nonzero term involving $z$ that appears
in the potential expansion scales like $x z^2$. In practice, this term
change the equilibrium shear itself, making it $z$--dependent. For a
barotropic gas as considered here, this violates the results that
rotation should be on cylinders \citep{tassoul78}, unless the
equilibrium profile of the density depends on $x$ as well as on
$z$. This is a significant modification to the standard
shearing box model. In practice, smoothing can only be achieved with
higher order terms that are in $(H/R)^2$, among which a term in $-z^4$
that results in a finite potential barrier out of which to climb,
albeit with an incorrect value. These complications were not addressed
in the SMI10 analysis, which makes the interpretation of their results
unclear.


Clearly, the discussion above shows that changing the vertical profile
of the tidal potential is a risky avenue, much beyond the scope of
this paper. It has to be done in a way that is consistent with all
the approximations involved in the shearing box. The extension of the
shearing-box model in that direction is promising for the analysis of outflow-related issues and should be the subject of future work.

\section{Conclusions and perspectives}
\label{conclusion_sec}

In this paper, we have analysed in detail the structure of the flow in
an accretion disk threaded by a weak vertical magnetic field. In
agreement with previous results (SI09, SMI10), we find that a strong
outflow develops, with mass-loss rates that deplete the gas content of
the computational box within a hundred dynamical times. However,
we found these mass-loss rates to depend strongly on the numerical
setup, thus preventing a reliable estimate of such outflow rates from being
made in shearing-box simulations. Nevertheless, we found that
the flow properties in the disk atmosphere exhibit robust features:
outflows are magneto-centrifugally accelerated, yet thermal pressure
has a non negligible effect.
We recover
the same invariants that characterized classical theories of
steady-state disk winds \citep[BP82;][]{pelletier&pudritz92}. We
stress however, that there are a number of important differences with such
theories: first, the outflows produced in the simulations are strongly
time-dependent, which means steady state models should be used with care. 
Second, the vertical outflow velocity is never seen to exceed the fast
magnetosonic speed. In addition, the positions of the slow
magnetosonic point and the Alfv\'en point are seen to depend on box
size. This highlights the fact  
that information can come back down the outflow, although we have not
been able to find any significant modifications of its properties by
changing the nature of the vertical boundary conditions. Finally, and maybe most
importantly, we found that the dynamical effect of these outflows on
accretion disks remains to be demonstrated: we investigated the ratio
of the angular momentum extraction by such outflows and
its radial transport by turbulence. These quantities might become
comparable for thin disks for which the escape velocity is larger by
more than one order of magnitude than the velocities observed in the
simulations. Thus, the ultimate fate of these outflows, either
successful launching or falling back onto the disk, remains unclear.

These results should be contrasted with a case of stronger, MRI stable magnetic
field. In that situation, \citet{ogilvie12} has argued, with reference
to previous work, that the local approximation can be used to
determine the mass-loss rate per unit area in magnetized outflows that
are accelerated along inclined poloidal magnetic field lines
\citep{blandford&payne82}. In turbulent disks as studied in the
present paper, this result does not hold. The occurrence of
transient, and presumably local, outflows appears to be a natural
consequence of the nonlinear development of the MRI in a stratified
disk in the presence of a net vertical magnetic field.  This
behaviour is different from that in incompressible disks, where no
vertical motion arises from nonlinear channel modes because the
vertical forces are cancelled by pressure gradients
\citep{goodman&xu94}.  Nevertheless, the connection between these
transient and 
local outflows and the large-scale jets observed from astrophysical
discs is still unclear.  Traditional models for large-scale outflows
have strong, ordered magnetic fields and Alfv\'en points high above
the surface of the disk, rather different from the situation in our
simulations.


In the light of our results, it appears that the shearing-box model is not
well adapted to answer these questions, unless new developments are
undertaken that go much beyond its original formulation. Both
quantitative mass-loss rates and the ultimate fate of these outflows may
have to be determined using global numerical simulations. This is a
formidable task, the computational cost of which is at the limit of
currently available facilities. But perhaps the good news of our work
is that moderate resolution and runtime are sufficient to
capture the properties of disk outflows. This gives hope that the
questions of the fate, properties and consequences of outflows for
accretion disk dynamics can be answered in the next few years.



\section*{ACKNOWLEDGMENTS}

We acknowledge the referee, D.Uzdensky, for a very detailed report than
significantly improved the paper. SF acknowledges funding from the
European Research Council under the European Union's Seventh Framework
Programme (FP7/2007-2013) / ERC Grant agreement n° 258729. GIO and HL
acknowledge support via STFC grant ST/G002584/1. GL acknowledges
support by the European Community via contract
PCIG09-GA-2011-294110. The simulations presented in this paper were
granted access to the HPC resources of CCRT under the allocation
x2012042231 made by GENCI (Grand Equipement National de Calcul
Intensif).

\bibliographystyle{aa}
\bibliography{author}

\appendix
\section{Linear modes in viscous and resistive stratified disks}

\begin{figure}
\begin{center}
\includegraphics[scale=0.29]{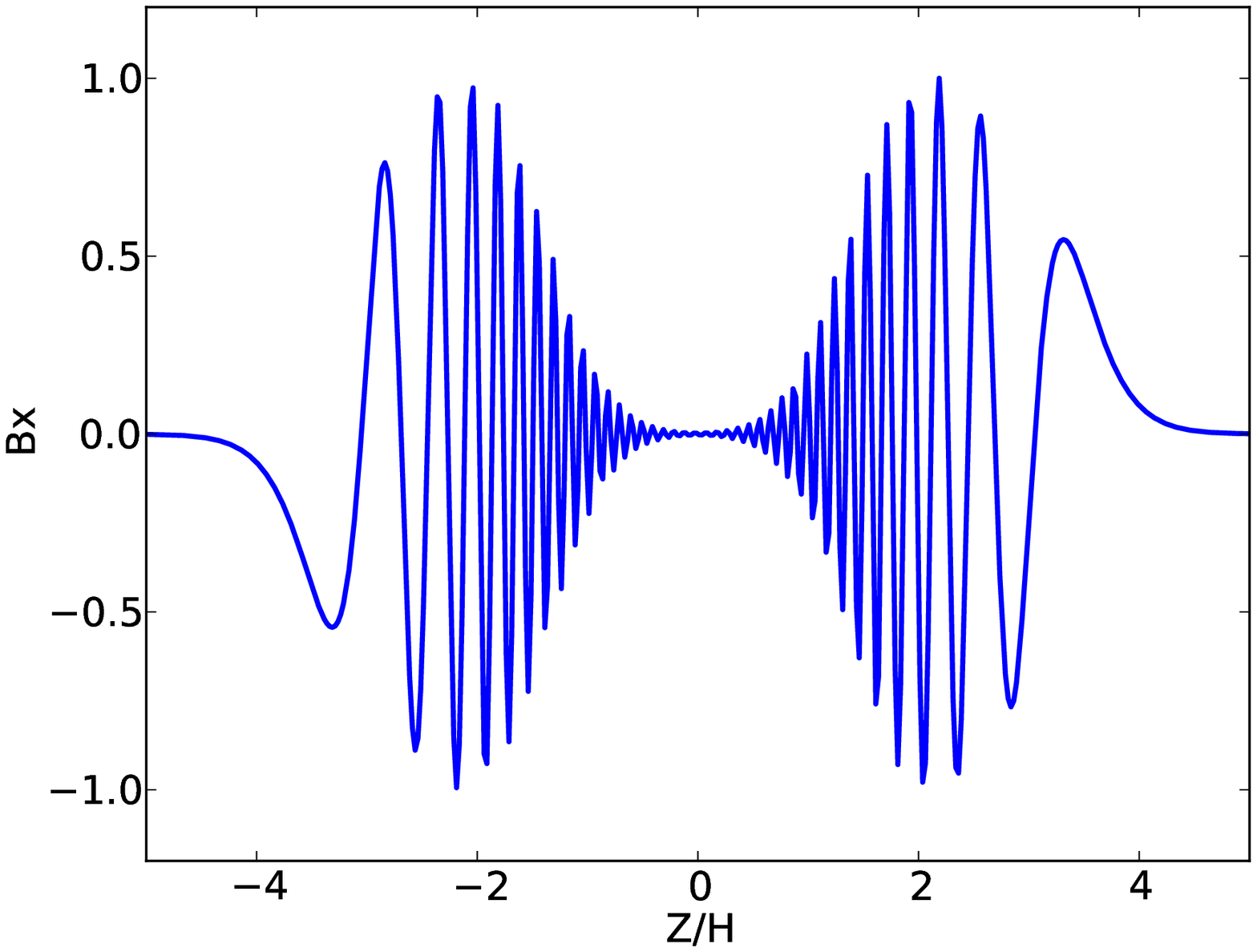}
\includegraphics[scale=0.29]{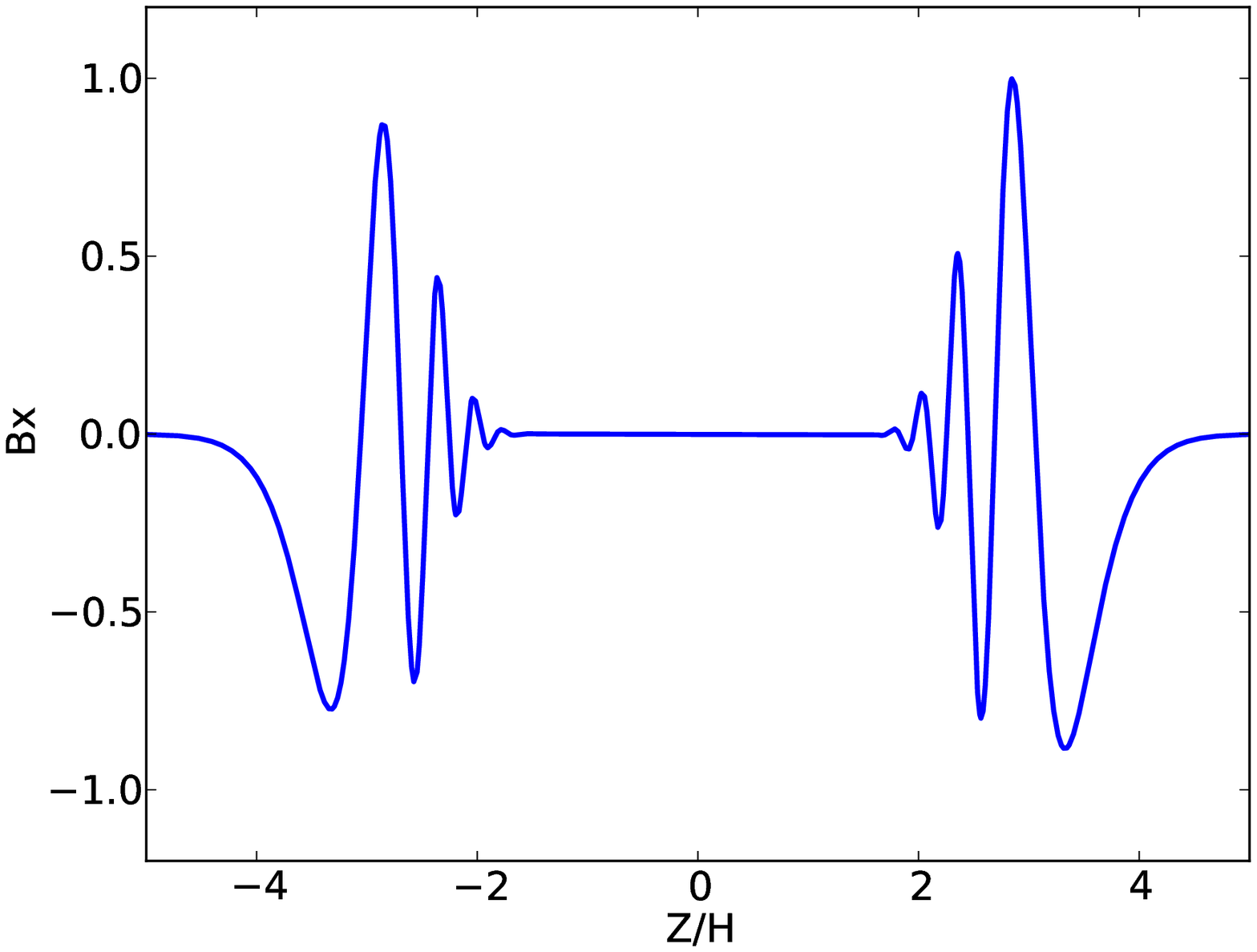}
\includegraphics[scale=0.29]{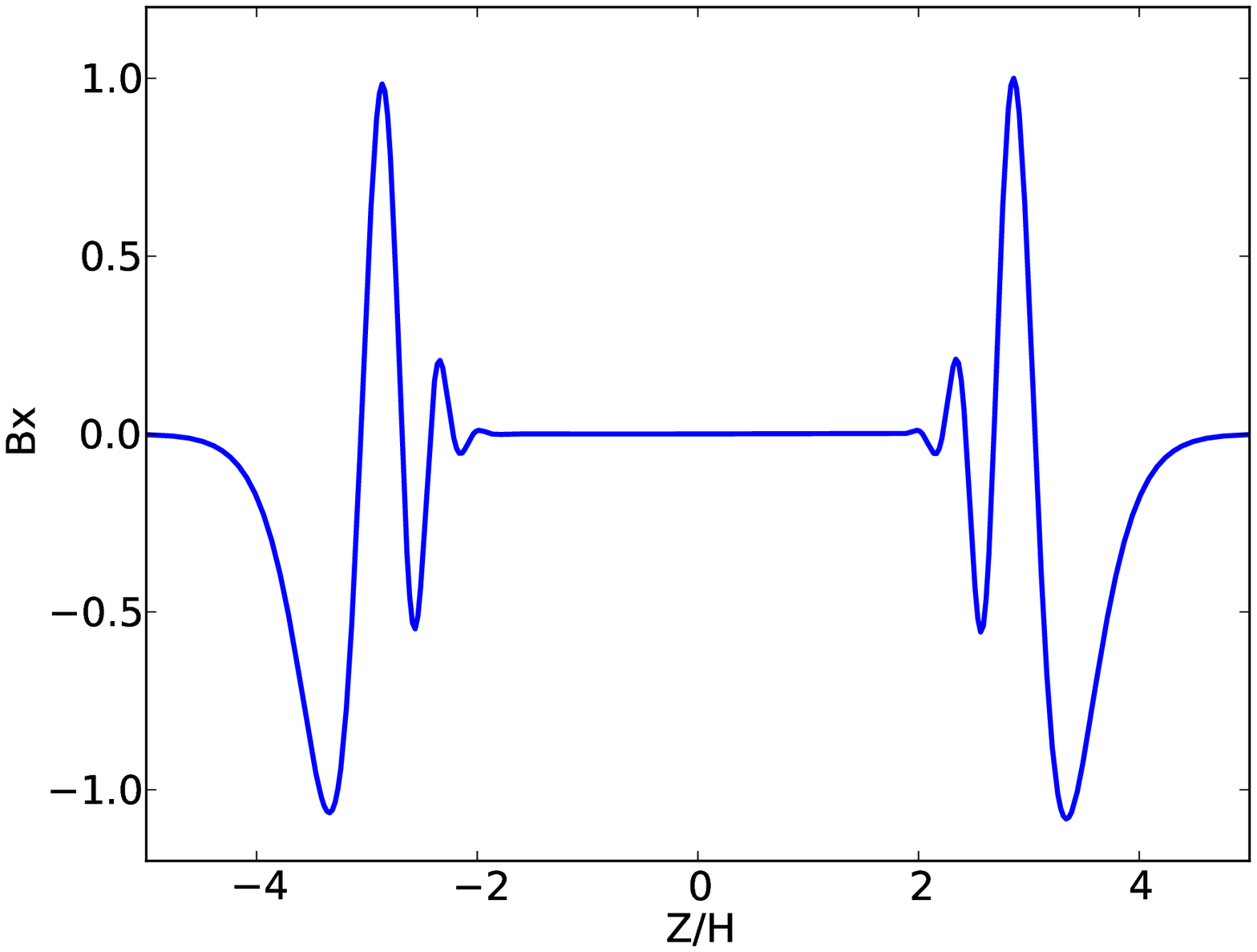}
\caption{Vertical profile of the radial magnetic field eigenmode for
three different sets of dissipation coefficients. From left to right,
$Rm=Re=10^6$, $Rm=Re=10^4$ and $Rm=Re=3000$.}
\label{eigenmode_fig}
\end{center}
\end{figure}

Consider an MRI channel flow of the following form:
\begin{align}
&\bb{u}^\text{ch} = H\Omega\,e^{st}\left[ \bb{i}\,u_x(z) +
 \bb{k}\,u_y(z)\right],\\ &\bb{B}^\text{ch} =
B_0\,e^{st}\left[\bb{i}\,B_x(z) + \bb{k}\,B_y(z) \right], \end{align}
where $H$ is the disk semithickness, $B_0$ is the amplitude of the
background vertical field, $s$ is the growth rate, and $u_x$, $u_y$,
$B_x$ and $B_y$ are dimensionless functions to be determined. Note
that in the limit of high midplane $\beta$ (anelastic approximation) 
this flow is both a linear and nonlinear solution to
the equations of non-ideal MHD in the vertically stratified shearing
box \citep[cf.][]{latteretal10}.

We denote the background density as $\rho=\rho_0\,h(z)$ where $h$ is a
dimensionless function, and then scale time by $1/\Omega$ and space by
$H$. The four equations to be solved are then \begin{align} s\,u_x &=
  2\,u_y +\frac{2}{\beta}\,\frac{1}{h}\,B_x'
  +\frac{1}{Re}\frac{1}{h}\,(h\,u_x')', \\ s\,u_y &= -\frac{1}{2}\,u_x
  +
  \frac{2}{\beta}\,\frac{1}{h}\,B_y'+\frac{1}{Re}\frac{1}{h}\,(h\,u_y')',
  \\ s\,B_x &= u_x' + \frac{1}{Rm}\,B_x'', \\ s\,B_y &= u_y' -
  \frac{3}{2}\,B_x + \frac{1}{Rm} B_y'', \end{align} where a prime
indicates differentiation with respect to $z$. This set of equations
can be solved 
via a pseudo-spectral method using Whittaker cardinal functions, as
outlined in Latter et al.~(2010). 

 Examples of these calculations are
plotted in Fig.~\ref{eigenmode_fig}. Note in these
profiles how the combined action of magnetic diffusion and viscosity
dramatically alters the mode morphology even for relatively large 
$Re$ and $Rm$. In the ideal case, the modes
are concentrated near the miplane, where they also exhibit their
shortest scales. These short scales, however, are
vulnerable to even a small amount of dissipation, and when
present the modes as a consequence tend
to localise near $z=\pm 2$-$3H$ where their characteristic scale is longer.

\end{document}